\documentclass[aps,reprint,amsmath,amssymb,superscriptaddress,showpacs,prb]%
{revtex4-2}
\usepackage{graphicx}
\usepackage{color}
\usepackage{lipsum}
\usepackage{dcolumn}
\usepackage{bm}

\renewcommand{\vec}{\mathbf}

\renewcommand{\r}{\vec{r}}
\renewcommand{\k}{\vec{k}}
\newcommand{\q}{\vec{q}}
\renewcommand{\d}{\mathrm{d}}

\newcommand{\rmi}{\text{i}}

\newcommand{\veczero}{\bm{0}}

\begin{document}

\title{Quantum states in disordered media. II. Spatial charge carrier distribution}

\author{A.~V.~Nenashev}
\thanks{On leave of absence from Rzhanov Institute of Semiconductor Physics
  and the Novosibirsk State University, Russia}
\email{nenashev\_isp@mail.ru}
\affiliation{Department of Physics and Material Sciences Center,
  Philipps-Universit\"at Marburg, D-35032 Marburg, Germany}

\author {S.~D.~Baranovskii}
\email{sergei.baranovski@physik.uni-marburg.de}
\affiliation{Department of Physics and Material Sciences Center,
  Philipps-Universit\"at Marburg, D-35032 Marburg, Germany}
\affiliation{Department f\"ur Chemie, Universit\"at zu K\"oln,
  Luxemburger Stra\ss e 116,
  50939 K\"{o}ln, Germany}

 \author{K.~Meerholz}
\affiliation{Department f\"ur Chemie, Universit\"at zu K\"oln,
  Luxemburger Stra\ss e 116,
  50939 K\"{o}ln, Germany}

%
%

\author{F.~Gebhard}
\email{florian.gebhard@physik.uni-marburg.de}
\affiliation{Department of Physics and Material Sciences Center,
Philipps-Universit\"at Marburg, D-35032 Marburg, Germany}

\date{\today}


\begin{abstract}
The space- and temperature-dependent electron distribution $n(\r,T)$
is essential for the theoretical description of the opto-electronic properties
of disordered semiconductors.
We present two powerful techniques to access $n(\r,T)$
without solving the Schr\"odinger equation.
First, we derive the density for non-degenerate electrons
by applying the Hamiltonian recursively to random wave functions (RWF).
Second, we obtain a
temperature-dependent effective potential from the application
of a universal low-pass filter (ULF) to the random potential
acting on the charge carriers in disordered media.
Thereby, the full quantum-mechanical problem is reduced to
the quasi-classical description of $n(\r,T)$ in an effective potential.
We numerically verify both approaches by comparison
with the exact quantum-mechanical solution.
Both approaches prove superior to the widely used localization landscape theory (LLT)
when we compare our approximate results for the charge carrier density and mobility
at elevated temperatures
obtained by RWF, ULF, and LLT with those from the exact
solution of the Schr\"odinger equation.
\end{abstract}
\maketitle   

\section{Introduction}
\label{introduction}

Short-range disorder potentials govern the opto-elec\-tron\-ic properties of a wide
variety of disordered materials
designed for applications in modern electronics~\cite{Baranovskii2006Book}.
In this work we address the theoretical description of the energetically low-lying,
spatially localized states
in several classes of disordered materials.

One broad class of widely studied disordered materials
are semiconductor alloys with electronic states
localized by a short-range disorder potential.
Alloys are crystalline semiconductors, such as A$_{\bar{x}}$B$_{1-\bar{x}}$,
whose lattice sites are occupied in a given proportion $\bar{x}$ by chemically
different isoelectronic atoms, $A$ and $B$.
Research on semiconductor alloys currently experiences a renaissance because
alloying is one of the most efficient tools to adjust material properties for
device applications. For example, alloying permits to tune lattice constants, effective
masses of charge carriers and, most importantly, the band-structure of the underlying
semiconductor.
In particular, band gaps in alloy semiconductors are sensitive
to the mole fractions~$\bar{x}$ and $1-\bar{x}$
of the alloy components. Since the band gap is a key property responsible for light
absorption and emission, band-gap tunability in a wide energy range opens
rich perspectives for applications of alloy semiconductors in various opto-electronic
devices.
Over recent years, this band-gap engineering has been applied
to nitride semiconductors used in modern
LEDs~\cite{WeisbuchNakamura2021}, to perovskites for applications
in photovoltaics~\cite{Hu2019,Paulina2022},
and to two-dimensional systems, such as transition metal dichalcogenides,
desired to miniaturize the corresponding devices toward nearly atomically
thin dimensions~\cite{Chaves2020,Masenda2021}.

Another class of disordered materials with short-range disorder potentials are
amorphous oxide semiconductors,
such as InGaZnO, desired for applications in thin-film transistors for transparent
and flexible
flat-panel displays~\cite{Nomura2004,IGZO2019}.
Also traditional amorphous semiconductors,
such as  hydrogenated amorphous silicon, hydrogenated amorphous carbon,
poly-crystalline and micro-crystalline silicon, belong to disordered materials,
whose opto-electronic properties are governed by short-range disorder potentials.
Such amorphous semiconductors are
desired for applications in Schottky barrier diodes, p–i–n diodes,
thin-film transistors, and thin-film solar cells~\cite{Baranovskii2006Book}.

Besides their unique properties favorable for device applications,
materials with short-range disorder
are advantageous systems for developing and testing theoretical descriptions
of disorder effects.
The compositional fluctuations in alloys
and structural short-range fluctuations in oxide and amorphous semiconductors cause
spatial fluctuations of the band gap which creates
a random potential acting on electrons and holes.
The isoelectronic substitution of the alloy components and/or amorphous atomic
structure creates a short-range fluctuating crystal potential, and the
physics is not complicated by long-range effects.
The random potential leads to the spatial localization of
charge carriers in electron states at low energies that
dominate the opto-electronic properties of disordered semiconductors.
Therefore, the low-energy localized electron states require an appropriate
theoretical description.

Already in the 1960s,
a powerful theoretical tool to calculate the density of low-energy localized states (DOS)
without solving the Schrödinger equation was developed by Halperin
and Lax~\cite{Halperin1966}.
They recognized that the spatial spread of wave functions
even in the low-energy localized states is much broader than the spatial scale
of the fluctuations in the random potential $V(\r)$. Hence, they proposed to
apply a filter to smooth $V(\r)$ where
the absolute square of the wave function serves
as their filter function~\cite{Halperin1966}. Later,
Baranovskii and Efros~\cite{Baranovskii1978} addressed the same problem
by a slightly different variational technique and confirmed the
result of Halperin and Lax for the low-energy tail of the DOS.
Theoretical studies
have been performed for various kinds of disordered semiconductors
that focus on the averaged DOS,
leaving aside the temperature-dependent spatial distribution
of the electron density $n(\r,T)$.
However, the knowledge of $n(\r,T)$ in a random potential is required %
to study theoretically charge transport and light absorption/emission
in disordered semiconductors.

The distribution $n(\r,T)$ is obtained from
the solution of the Schrödinger equation in the presence of a disorder potential.
However, solving the Schrödinger equation
is extremely demanding with respect to computational time and computer memory.
It is hardly affordable for applications to realistically large chemically
complex systems.
Therefore, theoretical
tools to obtain the essential features of localized
states without solving the Schr\"odinger equation are highly desirable.

One of such tools is the recently introduced `localization landscape
theory' (LLT)~\cite{Arnold2016,LL1_2017,LL2_2017,LL3_2017}.
The LLT is widely considered as one of the efficient theoretical approaches
to calculate the local density of states in disordered systems, for instance,
in light emitting
diodes~\cite{Chen2018APL,Vito2020,WeisbuchNakamura2021,Montoya2021,Tibaldi2021,Chaudhuri2021,Shen2021,Shen2022}
and in photodetectors~\cite{Tsai2020,Chow2020}. The LLT has been used to
simulate the carrier effective
potential fluctuations induced by alloy disorder in InGaN/GaN core-shell
microrods~\cite{Liu2019}. Furthermore, the LLT has been applied to compute
the eigenstate localization length at very low energies in a two-dimensional disorder
potential~\cite{Shamailov2021}. The LLT is sometimes considered capable
to reveal errors in the finite element method software in applications to
alloys~\cite{Zhan2021}. The LLT is an important ingredient
for calculating quantum corrections in drift-diffusion
models~\cite{Bertazzi2020,Donovan2021,Donovan2022}.
It has been used for the computation of light absorption in three-dimensional
InGaN alloys of different compositions~\cite{Banon2022}
and in mixed halide perovskites~\cite{Yun_Urbach2022}.

The LLT is based on a mathematical
theory of quantum localization that introduces an effective
localization potential.
Using the effective potential, the quantum
mechanical problem reduces to a quasi-classical description
of charge carriers localized in the effective potential landscape.
This allows the prediction of localization regions for electrons and holes,
of their corresponding energy levels, and of the (local) densities of states.
The effective potential can be directly implemented in a drift-diffusion model
of carrier transport
and into the calculation of absorption/emission transitions
so that physical features on a mesoscopic scale become accessible.

In the preceding paper~\cite{Gebhard}, it was shown that the LLT
becomes equivalent to the low-pass filter
approach of Halperin and Lax~\cite{Halperin1966}
for the specific case of a Lorentzian filter when applied to the Schr\"odinger equation
with a constant mass.
In this work, we propose
two recipes for calculating $n(\r,T)$ in disordered systems
without solving
the Schr\"odinger equation, both of them prove to be superior to the LLT.

The first recipe is the random-wave-function (RWF) algorithm described
in Sect.~\ref{subsec:synthesis}.
In this approach, many sets of random wave functions are generated
for a given realization
of the potential $V(\r)$
acting on the charge carriers. Such procedure has been suggested by Lu and Steinerberger \cite{Lu2018} in order to search for the low-lying eigenfunctions of various linear operators.
After a repeated application of the thermal operator,
the temperature-dependent electron density
is found from averaging the results over different sets.
This procedure provides reliable results for $n(\r,T)$
for fairly large systems.
The RWF approach has the following advantages over the LLT.
While the output of the LLT depends on the choice of an adjustable parameter
that can be found only by comparison with the exact solution,
the RWF scheme does not contain any adjustable parameters.
Furthermore, the accuracy of the RWF approach for computing $n(\r,T)$
can be improved systematically,
whereas the accuracy of the LLT is limited by construction.
Moreover, all temperatures are accessible for the RWF algorithm,
whereas LLT is valid only at sufficiently high $T$,
as discussed in Sect.~\ref{sec:ComparisonwithLLT}.

The second recipe employs an effective temperature-dependent potential $W(\r,T)$
that permits a quasi-clas\-si\-cal calculation of the particle density.
The effective potential is obtained from the random potential $V(\r)$ by applying
a universal low-pass filter (ULF),
as described in Sect.~\ref{sec:low_pass_filter}.
This approach resembles the one suggested by Halperin and Lax~\cite{Halperin1966}.
In our work, we use a low-pass filter
to compute the spatial electron distribution $n(\r,T)$
while Halperin and Lax and their successors
focused on the features of the averaged DOS.
The ULF algorithm is equivalent to the LLT inasmuch it replaces the random
potential $V$  by an effective potential $W$ that
is used for a quasi-classical description of the electron distribution.
However, the ULF scheme of converting $V$ into $W$ principally
differs from the LLT scheme.
While LLT is based on solving a system of algebraic equations,
the ULF algorithm employs Fast-Fourier Transformation to calculate
the effective potential $W$
so that very large systems can be addressed. %
More importantly, the ULF description leads to an explicit
temperature dependence of the effective potential, $W\equiv W(\r,T)$.
On the contrary, the effective potential in LLT does not depend on~$T$.
This feature is the main disadvantage of the LLT as compared to the ULF algorithm.
Consequently, the range of applicability for the ULF scheme is much
broader than that for the LLT, as illustrated in Sect.~\ref{sec:ComparisonwithLLT}.
Typically, the ULF description
encompasses the experimentally accessible temperature regime.

In our theoretical consideration, we consider electrons as non-interacting particles. Commonly, electron-electron interactions are taken into account in the self-consistent solution of the Schr\"odinger equation and Poisson equation. Since we suggest to replace the solution of the Schr\"odinger equation in search for $n(\r,T)$ by RWF or ULF techniques, the latter methods should be exploited in a self-cosistent combination with the Poisson equation in order to simulate realistic conditions met in devices.

Our paper is structured as follows. In Sect.~\ref{sec:density-calc},
we develop the RWF algorithm to calculate
$n(\r,T)$ for dilute charge carriers in disordered media.

In Sect.~\ref{sec:eff-potential},
we derive the temperature-dependent
effective potential $W(\r,T)$ that permits a semi-classical description
of the particle density.
The algorithm employs a universal low-pass filter (ULF) function
and a temperature-dependent shift that we derive and analyze in detail.

The space- and temperature-dependent electron distribution $n(\r, T)$ is the key
ingredient for the theoretical treatment of charge transport.
In Sect.~\ref{sec:ComparisonwithLLT}, we analyze the results of
different approaches for the particle density and
for the charge carrier mobility in disordered media at elevated temperatures.

Conclusions in Sect.~\ref{sec:conclusions} close the main part of our presentation.
Minor technical aspects of our work are deferred to two appendices.

\section{Electron density in disordered media}
\label{sec:density-calc}

In this section we develop the random-wave-function al\-go\-rithm
to calculate the electron density in non-de\-gen\-er\-ate systems at finite temperatures.
We start with definitions in Sect.~\ref{subsec:nofrandT}.
Then, in Sect.~\ref{subsec:analysis}, we express the particle density
as an average over a random combination of eigenstates.
In Sect.~\ref{subsec:synthesis}, we summarize the
main steps of the algorithm.
Finally, in Sect.~\ref{subsec:examples},
we illustrate the algorithm and demonstrate its efficacy
for disordered systems in one, two, and three dimensions.

\subsection{Electron density for potential problems}
\label{subsec:nofrandT}

Let us consider a system of non-interacting electrons
governed by a single-particle Hamiltonian
\begin{equation}
  \hat{H}=\hat{T}+\hat{V} \; ,
  \end{equation}
where $\hat{T}$ describes their kinetic energy and $\hat{V}$
some potential in the specimen volume $\Omega$.
We are interested to calculate
the electron density $n(\r,T)$ in thermal equilibrium at temperature~$T$.
The eigenstates of the Hamiltonian $|\varphi_i\rangle$ with eigenenergies $\epsilon_i$,
\begin{equation}
  \hat{H}|\varphi_i\rangle = \epsilon_i |\varphi_i\rangle
  \label{eq:Schroedinger}
  \end{equation}
have the position-space representation $\varphi_i(\r)$,
\begin{equation}
\varphi_i(\r) =   \langle \r |\varphi_i\rangle  \;,
  \end{equation}
and $|\varphi_i(\r)|^2$ gives the probability to find the electron at $\r$ in
the eigenstate $|\varphi_i\rangle$.
Correspondingly, the eigenfunctions are normalized by unity,
\begin{equation}
  \int_{\Omega} |\varphi_i(\r)|^2 \, \d\r = 1
  \label{eq:phi-norm}
\end{equation}
because the electron can be found somewhere in the specimen.

In the presence of a finite number of electrons~$N$ in~$\Omega$,
the electron density $n(\r,T)$  at temperature~$T$ is defined by
\begin{equation}
  n(\r,T) = 2 \sum_i |\varphi_i(\r)|^2 f_{\rm FD}(\epsilon_i) \; ,
  \label{eq:n-def}
\end{equation}
where the factor two accounts for the spin degeneracy and
$f_{\rm FD}(\epsilon)$ is the equilibrium (Fermi-Dirac)
distribution function,
\begin{equation}
f_{\rm FD}(\epsilon)=\frac{1}{1+e^{\beta(\epsilon-\mu)}} \; ,
\end{equation}
where $\beta=1/(k_{\rm B} T)$ with the Boltzmann constant $k_{\rm B}$
and $\mu\equiv \mu(T)$ is the chemical potential.
It is fixed by the condition
\begin{equation}
n= \frac{N}{\Omega} = \int_{\Omega} \d\r \, n(\r,T) \; .
\end{equation}
In most disordered materials, the average electron density $n(\r,T)$ is small
so that the electrons' Fermi statistics can be ignored
(non-degenerate case). Then, the Fermi-Dirac distribution can be replaced by
the Maxwell-Boltzmann distribution
\begin{equation}
  f_{\rm MB}(\epsilon) = e^{-\beta(\epsilon - \mu)} \; .
  \label{eq:f-Maxwell-Boltzmann}
\end{equation}
In the Maxwell-Boltzmann approximation we can factor out the chemical potential,
\begin{equation}
  n(\r,T) = e^{\beta\mu} \; \tilde n(\r,T) \; ,
  \label{eq:n-via-tilde-n}
\end{equation}
where the reduced electron density $\tilde{n}(\r,T)$ is
determined only by the eigenstates $\varphi_i(\r)$,
their eigenenergies $\epsilon_i$,
and the temperature,
\begin{equation}
  \tilde{n}(\r,T) = 2 \sum_i |\varphi_i(\r)|^2 \exp(-\beta\epsilon_i) \; .
  \label{eq:tilde-n-def}
\end{equation}
We note in passing that the approximation of non-degenerate
electrons does not hold for band-structure theory
in solid-state physics. Therefore, our formalism below cannot be applied, e.g.,
in density-functional theory calculations.

For numerical calculations, we assume that the Hamiltonian $\hat{H}$
is defined on a regular grid with $L$ sites. The volume assigned to a grid node
thus is $\Delta V=\Omega/L$.
A wave function takes complex amplitudes
on nodes of the lattice.
The value $\psi(\r)$ of some state
$|\psi\rangle$ at $\r$ is obtained from
\begin{equation}
  \psi(\r) = \frac{1}{\sqrt{\Delta V}} \, \langle\r_n|\psi\rangle \; ,
  \label{eq:psi-Dirac}
\end{equation}
where $\r$ is in the grid-node volume $\Delta V$ around the grid point $\r_n$.

As a consequence of the discretization, the set of eigenenergies $\{ \epsilon_i \}$
is bounded within a finite range,
\begin{equation}
  \forall i: \epsilon_{\text{min}} \le \epsilon_i \le \epsilon_{\text{max}}
  \label{eq:epsilon-bounds}
\end{equation}
with some lower and upper bounds $\epsilon_{\text{min}}$
and $\epsilon_{\text{max}}$, respectively.
Generically,  $\epsilon_{\text{max}}$ is determined by the distance between grid points,
whereas $\epsilon_{\text{min}}$ is a defining property of the problem itself.
We make the discretization fine enough, $L\gg 1$, so that we
can safely assume that
\begin{equation}
  |\epsilon_{\text{max}}| \gg |\epsilon_{\text{min}}| \; .
  \label{eq:epsilon-max-much-greater}
\end{equation}
For simplicity, we assume that the Hamiltonian $\hat{H}$ can be represented
by a real symmetric matrix. To simplify the notations,
we also assume that its eigenfunctions $\varphi_i(\r)$
are real. The generalization to complex-valued Hamiltonian matrices
and eigenfunctions is straightforward.

\subsection{Analysis}
\label{subsec:analysis}

Consider a set $\{ c_i^{(0)} \}$ of $L$ independent random variables,
one per eigenstate $|\varepsilon_i\rangle$.
Each variable $c_i^{(0)}$
is taken from a normal distribution with expectation value zero and variance unity.
Thus, when we take a large number $N_{\rm R}$ of realizations~${\rm R}$
of the sets we obtain
\begin{eqnarray}
  \langle c_{i,R}^{(0)} \rangle_{\rm R} &=& \frac{1}{N_{\rm R}}
  \sum_{R=1}^{N_{\rm R}} c_{i,R}^{(0)} =0   \; ,\nonumber \\
	\langle c_{i,R}^{(0)} c_{j,R}^{(0)}
        \rangle_{\rm R} &=& \frac{1}{N_{\rm R}} \sum_{R=1}^{N_{\rm R}}
        c_{i,R}^{(0)} c_{j,R}^{(0)}
        = \delta_{i,j} \; .
         \label{eq:c0-statistics}
\end{eqnarray}
For a given realization~${\rm R}$, we use the random variables $c_{i,R}^{(0)}$
as coefficients to construct a wave-function $\psi_{\rm R}^{(0)}(\r)$
\begin{equation}
  \psi_{\rm R}^{(0)}(\r) = \sum_{i=1}^L c_{i,R}^{(0)}\, \varphi_i(\r) \; .
  \label{eq:psi0-def}
\end{equation}
Obviously, the mean value of these wave-functions over many realizations vanishes,
\begin{equation}
	\langle \psi_{\rm R}^{(0)}(\r) \rangle_{\rm R} =
	\sum_{i=1}^L \langle c_{i,R}^{(0)} \rangle_{\rm R} \varphi_i(\r) = 0 \; .
        \label{eq:psi0-mean}
\end{equation}
For the calculation of the two-point correlation, we assume that $\r_n$ and $\r_m$
are grid nodes to write
\begin{eqnarray}
	\langle \psi_{\rm R}^{(0)}(\r_n) \psi_{\rm R}^{(0)}(\r_m) \rangle_{\rm R}
	&=& \sum_{i,j} \langle c_{i,R}^{(0)} c_{j,R}^{(0)} \rangle_{\rm R}
	\varphi_i(\r_n) \varphi_j(\r_m) \nonumber \\
	&=& \sum_i \varphi_i(\r_n) \varphi_i(\r_m) \nonumber \\
&=& \frac{1}{\Delta V} \sum_i \langle\r_n|\varphi_i\rangle \langle\varphi_i|\r_m\rangle
        \nonumber \\
	&=& \frac{1}{\Delta V} \langle\r_n|\r_m\rangle \nonumber \\
	&=& \begin{cases}
		1/\Delta V & \text{if $n= m$}\; , \\
		0          & \text{otherwise} \; .
	\end{cases}
        \label{eq:psi0-correlators}
\end{eqnarray}
Here we used eqs.~(\ref{eq:psi-Dirac}), (\ref{eq:c0-statistics}), (\ref{eq:psi0-def}),
and the fact that the eigenfunctions $|\varphi_i\rangle$ form a complete orthonormal set.
Eq.~(\ref{eq:psi0-correlators}) shows that
the values of $\psi_{\rm R}^{(0)}(\r_n)$ at different grid nodes are independent random
variables that obey a normal distribution with expectation value equal to zero
and variance equal to $1/\Delta V$.

To introduce the temperature into the problem,
we consider the wave function $\psi_{\rm R}(\r)$
that is related to $\psi_{\rm R}^{(0)}(\r)$ via
\begin{equation}
  |\psi_{\rm R}\rangle = e^{-\beta\hat{H}/2} \, |\psi_{\rm R}^{(0)}\rangle \; .
  \label{eq:psi-def}
\end{equation}
We expand it over the eigenfunctions~$\varphi_i(\r)$,
\begin{equation}
  \psi_{\rm R}(\r) = \sum_i c_{i,R} \varphi_i(\r)  \; ,
  \label{eq:psi-expansion}
\end{equation}
to find the relation between the two sets of coefficients $c_i$ and $c_i^{(0)}$.
To this end we apply the operator $e^{-\beta\hat{H}/2}$
to both sides of eq.~(\ref{eq:psi0-def})
and take into account that $|\varphi_i\rangle$ are
eigenstates of the Hamiltonian. This gives
\begin{equation}
  c_{i,R} = c_{i,R}^{(0)} \exp(-\beta\epsilon_i/2) \; .
  \label{eq:c-via-c0}
\end{equation}
Using eqs.~(\ref{eq:c0-statistics}) and~(\ref{eq:c-via-c0}), we thus find that
\begin{equation}
  \langle c_{i,R }c_{j,R} \rangle_{\rm R} = \delta_{i,j} \exp(-\beta\epsilon_i) \; .
  \label{eq:c-correlators}
\end{equation}
With the help of this correlation, we find the mean value of $|\psi_{\rm R}(\r)|^2$,
\begin{eqnarray}
	\langle |\psi_{\rm R}(\r)|^2  \rangle_{\rm R}
	&=& \sum_{i,j} \langle c_{i,R} c_{j,R} \rangle_{\rm R}
	\varphi_i(\r) \varphi_j(\r) \nonumber \\
	&=& \sum_i \exp(-\beta\epsilon_i) |\varphi_i(\r)|^2 \; ,
        \label{eq:psi2}
\end{eqnarray}
or, comparing with eq.~(\ref{eq:tilde-n-def}),
\begin{equation}
  \langle |\psi_{\rm R}(\r)|^2  \rangle_{\rm R} = \frac{1}{2}
  \tilde n(\r,T) \; .
  \label{eq:psi2-via-tilde-n}
\end{equation}
Hence, the electron density can be calculated by averaging
the functions $|\psi_{\rm R}(\r)|^2$ over
many realizations~${\rm R}$.
This is the basic idea of the random-wave-function algorithm that we present
in more detail in the next subsection.

Our algorithm requires the application of the exponential
operator $\exp(-\beta \hat{H})$
to a given wave function. In practical applications,
this cannot be done exactly. We use an algebraic approximation,
\begin{equation}
  e^{-\beta\hat H/2} \approx (1 - \alpha\hat H)^M \; ,
  \label{eq:exp-H-via-iterations}
\end{equation}
with a natural number $M$ equal to
\begin{equation}
  M \approx \frac{\beta}{2\alpha} \; .
  \label{eq:M-def}
\end{equation}
The parameter $\alpha$ can be chosen to minimize the residual error.
The action of operator $\exp(-\beta\hat{H}/2)$ suppresses
the high-energy components in $|\psi_{\rm R}\rangle$ relative to its low-energy ones.
The algebraic approximation on the right-hand side
of eq.~(\ref{eq:exp-H-via-iterations}) must have a similar effect,
i.e., the factor $(1 - \alpha\epsilon_{\text{max}})^M$
has to be small with respect to $(1 - \alpha\epsilon_{\text{min}})^M$
in absolute values, thence
\begin{equation}
  |1 - \alpha \epsilon_{\text{max}}| < |1 - \alpha \epsilon_{\text{min}}|
  \label{eq:alpha-e-min-e-max}
\end{equation}
must hold. The term $\alpha \epsilon_{\text{min}}$ can be neglected
because $|\epsilon_{\text{min}}| \ll |\epsilon_{\text{max}}|$
due to eq.~(\ref{eq:epsilon-max-much-greater}).
Then, the inequality~(\ref{eq:alpha-e-min-e-max}) reduces to
$-1 < 1 - \alpha \epsilon_{\text{max}} < 1$.
We note that $\epsilon_{\text{max}} > 0$ because this quantity is determined mainly
by the maximal kinetic energy in a discretized system. Therefore,
we obtain the following constraints on $\alpha$,
\begin{equation}
  0 < \alpha < \frac{2}{\epsilon_{\text{max}}} \; .
  \label{eq:alpha-condition}
\end{equation}
On the other hand, a good approximation in eq.~(\ref{eq:exp-H-via-iterations})
requires that the number $M$ should be large.
Using eq.~(\ref{eq:M-def}) and recalling
that $\beta = 1/(k_{\rm B} T)$, we can rewrite the condition $M \gg 1$
in the form
\begin{equation}
  \alpha \ll \frac{1}{k_{\rm B} T} \; .
  \label{eq:alpha-condition-kT}
\end{equation}
Usually $k_{\rm B} T \ll \epsilon_{\text{max}}$,
and, as a consequence, the stronger condition~(\ref{eq:alpha-condition})
overrides the weaker condition~(\ref{eq:alpha-condition-kT}).

\subsection{Synthesis: the random-wave-function algorithm}
\label{subsec:synthesis}

The considerations presented in the previous subsection
suggest the following random-wave-function algorithm
for calculating the electron
density $n(\r,T)$ in non-degenerate non-interacting electron systems.

\begin{itemize}
  \item[S1] For each grid node $\r_n$, choose a random number $\psi_{\rm R}^{(0)}(\r_n)$
from a normal distribution with expectation value zero and variance $1/\Delta V$.
All these random numbers are chosen independently,
so that the function $\psi_{\rm R}^{(0)}(\r)$
itself represents a sample of Gaussian white noise.
The set of all  numbers constitutes the realization~${\rm R}$.

\item[S2] Apply the following transformation $M$ times,
\begin{equation}
  \underline{\psi}^{(m)} = \underline{\psi}^{(m-1)} - \alpha
  \underline{\underline{H}} \, \underline{\psi}^{(m-1)}
        \label{eq:step2}
\end{equation}
with $m = 1, 2, \ldots , M$. Here, $\underline{\psi}^{(m)}$ is the vector that
represents the wave function on the grid, and
$\underline{\underline{H}}$ is the matrix representation of the Hamiltonian
on the grid. Recall that $M$ is given by eq.~(\ref{eq:M-def}), and $\alpha$ obeys
the inequalities~(\ref{eq:alpha-condition}). In our calculations we find that
\begin{equation}
  \alpha = \frac{1.5}{\epsilon_{\text{max}}}
  \label{eq:step2-alpha-choice}
\end{equation}
is a suitable choice.
\item[S3] Calculate an estimate of the reduced electron density $\tilde{n}_{\rm R}(\r,T)$ as
\begin{equation}
  \tilde n_{\rm R}(\r_n,T) = 2 \, |\psi^{(M)}(\r_n)|^2
  \label{eq:step3}
\end{equation}
on each grid point $\r_n$.
\end{itemize}
The steps S1--S3 are carried out for a large number~$N_{\rm R}$ of
realizations~${\rm R}$ of $\psi_{\rm R}^{(0)}(\r)$.
Then, the electron density $n(\r,T)$ is the arithmetic mean
of the functions $\tilde{n}_{\rm R}(\r)$ obtained for different realizations R,
multiplied by a chemical-potential-related factor $e^{\beta\mu}$,
\begin{equation}
  n(\r,T) = e^{\beta\mu} \langle \tilde{n}_{\rm R} (\r,T) \rangle_{\rm R} \; .
  \label{eq:n-from-algorithm}
\end{equation}

The larger the number of realizations $N_{\rm R}$,
the more accurate is the calculated electron density $n(\r,T)$.
A convenient measure of the accuracy is the relative error $\Delta(\r,T)$,
\begin{equation}
  \Delta(\r,T) = \frac{ n^{\text{rwf}}(\r,T) - n^{\text{exact}}(\r,T) }{
    n^{\text{exact}}(\r,T) } \; ,
  \label{eq:Delta-def}
\end{equation}
where $n^{\text{rwf}}(\r,T)$
is calculated using the random-wave-function algorithm,
and the exact value $n^{\text{exact}}(\r,T)$
is defined by eqs.~(\ref{eq:n-via-tilde-n}) and~(\ref{eq:tilde-n-def}).
As we show in Appendix~\ref{sec:appendix-accuracy},
the expectation value of $\Delta^2$ is equal to
\begin{equation}
  \left\langle [\Delta(\r,T)]^2 \right\rangle = \frac{ 2 }{ N_{\rm R} } \; ,
  \label{eq:Delta2-expected}
\end{equation}
i.e., the typical deviation from the exact result is not larger than
$\sqrt{1/N_{\rm R}}$.

To calculate the mean value in eq.~(\ref{eq:n-from-algorithm}),
it is necessary to store the cumulative function
\begin{equation}
  \sum_{\rm R} \tilde n_{\rm R}(\r)
  \label{eq:sum-to-store}
\end{equation}
in the memory, and to update it after processing of each realization~${\rm R}$.

It is easy to estimate the resources of memory and computation time (or number
of operations) demanded by the random-wave-function algorithm.
It is enough to store in the memory $4L$ real numbers, namely,
$L$ entries for the potential energy in the Hamiltonian,
$2L$ numbers for the wave-functions
$\underline{\psi}^{(m)}$ and $\underline{\psi}^{(m-1)}$
at Step~S2, and $L$ values for the sum~(\ref{eq:sum-to-store}).
Almost all computation time is spent at Step~S2.
It takes ${\cal A}\simeq L M N_{\rm R}$ operations.
Indeed, applying the Hamiltonian
to a wave-function costs ${\cal O}(L)$ operations due to the fact
that the potential is local and the kinetic energy is a local second derivative.
Applying the Hamiltonian repeats $M$~times
for each of $N_{\rm R}$ realizations.
Let us further express the quantities $L$, $M$ and $N_{\rm R}$
via the sample volume $\Omega$, the distance between grid points $a$,
and the desired relative error~$\Delta$ using eqs.~(\ref{eq:M-def}),
(\ref{eq:step2-alpha-choice}) and~(\ref{eq:Delta2-expected}),
\begin{equation}
    L \simeq \frac{\Omega}{a^d} \, , \quad
  M \simeq \frac{\epsilon_{\text{max}}}{k_B T} \simeq \frac{\hbar^2}{m a^2 k_B T}\, ,
  \quad
  N_{\rm R} \simeq \frac{1}{\Delta^2}
  \label{eq:LMR-estimate-1}
\end{equation}
in a $d$-dimensional system. Therefore,
the estimate for the number of arithmetical operations becomes
\begin{equation}
  {\cal A}
\simeq L M N_{\rm R} \simeq \frac{\hbar^2}{m k_B T} \;
  \frac{\Omega}{a^{d+2} \Delta^2}
  \label{eq:LMR-estimate-2}
\end{equation}
when expressed in terms of physical quantities.
It is important to note that our algorithm is easily parallelized because
different realizations~${\rm R}$ of
the wave-function $\psi_{\rm R}^{(0)}(\r)$ can be processed
completely independently.

\subsection{Example: electrons in a random potential}
\label{subsec:examples}

In this subsection
we demonstrate how to use the random-wave-function algorithm.

\subsubsection{White-noise potential and dimensionless units}

To be definite, we consider a non-interacting electron gas in
a one-dimensional (1D), a two-dimensional (2D), and a three-dimensional (3D)
white-noise random potential. Such a potential $V(\r)$ is characterized by
the statistical properties
\begin{equation}
  \langle V_{\rm R} (\r) \rangle_{\rm R} = 0 , \qquad
  \langle V_{\rm R}(\r) \, V_{\rm R}(\r+\r') \rangle_{\rm R}
  = S \, \delta(\r') \; ,
	\label{eq:white-noise-def}
\end{equation}
when the average is taken over many realization~${\rm R}$ of the random
potential $V_{\rm R} (\r)$. The strength of the disorder is characterized
by the parameter~$S$.
The actual value of $S$ depends on the origin of the random potential.
For instance, compositional disorder in a semiconductor
alloy $A_{\bar{x}}B_{1-\bar{x}}$ leads to~\cite{Raikh1990}
\begin{equation}
	S = \frac{[\alpha(\bar{x})]^2 \bar{x} (1-\bar{x})}{N} \, ,
	\label{eq:S-compositional}
\end{equation}
where $N$ is the concentration of atoms
and $\alpha_c(\bar{x}) = \d E_c(\bar{x})/\d \bar{x}$ for electrons,
$\alpha_v(\bar{x}) = \d E_v(\bar{x})/\d \bar{x}$ for holes
as a function of the band edge positions $E_{c,v}(\bar{x})$.

We discretize this system using a regular grid of points, equally spaced along
each dimension. This is a quadratic grid in 2D,
and a simple cubic grid in 3D.
The distance between neighboring grid points is~$a$.
To discretize the single-particle Hamiltonian $\hat H$,
we use the simplest difference scheme for the Laplacian operator,
\begin{eqnarray}
  \hat{H}_{\rm R} \psi(\r) &=& -\frac{\hbar^2}{2m} \sum_{s=1}^d
  \frac{ \psi(\r-a\vec e_s) + \psi(\r+a\vec e_s) - 2\psi(\r) }{ a^2 }
  \nonumber \\
&&	+ V_{\rm R}(\r) \psi(\r) \; ,
	\label{eq:H-discretized}
\end{eqnarray}
where $\r\equiv \r_n$ is a grid point,
and $\vec e_1 , \ldots , \vec e_d$ are unit vectors along the coordinate axes.
The realization $V_{\rm R}(\r_n)$ of the random potential at the grid points are chosen
as independent random numbers, uniformly distributed within the range
\begin{equation}
	-\sqrt{\frac{3S}{a^d}} < V_{\rm R}(\r_n) < \sqrt{\frac{3S}{a^d}} \; .
	\label{eq:V-range}
\end{equation}
One can easily see that this distribution satisfies
the properties~(\ref{eq:white-noise-def}).
Indeed, the discretization converts the delta-function,
\begin{equation}
	\delta(\r-\r') \to
	\begin{cases}
		1/a^d & \text{ if } \r\equiv \r_n = \r_m \equiv \r' \; , \\
		0     & \text{ if } \r_n \neq \r_m \; ,
	\end{cases}
	\label{eq:delta-function-discretized}
\end{equation}
and therefore eq.~(\ref{eq:white-noise-def}) takes the form
$\langle V_{\rm R}(\r_n) \rangle_{\rm R} = 0$, 
$\langle [V_{\rm R}(\r_n)]^2 \rangle_{\rm R} = S/a^d$, and
$\langle V_{\rm R}(\r_n) \, V_{\rm R}(\r_m) \rangle_{\rm R} = 0$ for $\r_n \neq \r_m$.
The uniform distribution~(\ref{eq:V-range}) meets these criteria.

It is convenient to work with dimensionless quantities. We note that
it is always possible to fix the values of the Planck constant $\hbar$,
the Boltzmann constant $k_B$, the effective mass $m$,
and the disorder strength $S$ by a proper choice of physical units.
So we set all these quantities to unity,
\begin{equation}
	\hbar = k_B = m = S = 1 \; .
	\label{eq:dimensionless}
\end{equation}
To do this, we choose a physical unit of length
\begin{equation}
\ell_0=	\left( \frac{\hbar^4}{m^2 S} \right)^{1/(4-d)} \; ,
	\label{eq:length-unit}
\end{equation}
and the corresponding unit of energy
\begin{equation}
\varepsilon_0	= \left( \frac{m^d S^2}{\hbar^{2d}} \right)^{1/(4-d)} \; .
	\label{eq:energy-unit}
\end{equation}
Then, the unit of time is $t_0=\hbar/\varepsilon_0$, and the unit of temperature
is $T_0=\varepsilon_0/k_{\rm B}$.
As an example, we mention that in a 2D system with compositional disorder,
for $\alpha = 0.5\, \text{eV}$,
$N = 10^{15}\, \text{cm}^{-2}$, $x = 0.5$, and $m=0.1 \, m_0$,
the units of length, energy, and temperature are equal to
$\ell_0=9.6\, \text{nm}$, $\varepsilon_0=8.3\,\text{meV}$, and
$T_0=96\, \text{K}$, respectively.

In the rest of this section, the physical units are chosen such
that the equalities~(\ref{eq:dimensionless}) hold.
The dimensionless value of the discretization step
is set to $a=0.1\ell_0$,
and the dimensionless value of temperature is set to $T=T_0\equiv 1$,
unless stated otherwise.
Periodical boundary conditions apply.

\subsubsection{One-dimensional case}

\begin{figure}[t]
	\includegraphics[width=\linewidth]{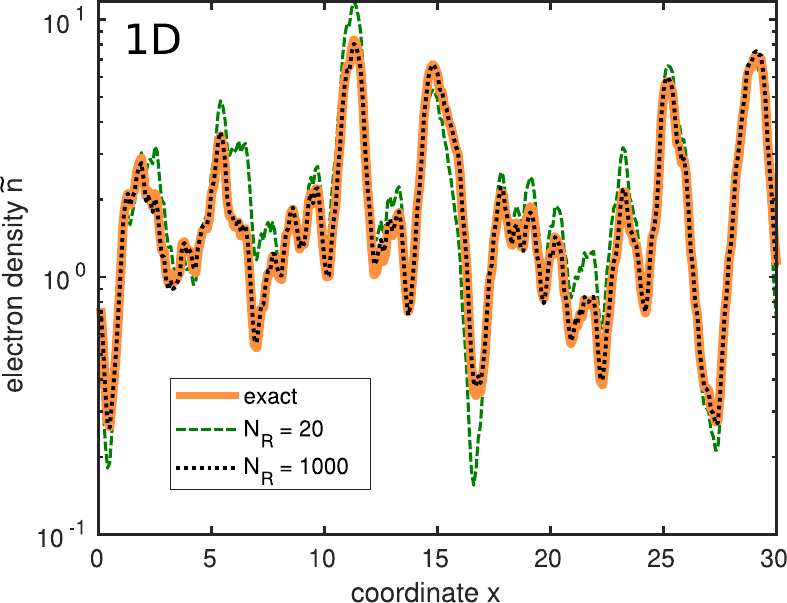}
	\caption{Reduced electron density $\tilde{n}(x,T) = e^{-\beta\mu} n(x,T)$
          as a function of coordinate $x$ in a sample with one-dimensional
          white-noise potential.
          Solid orange line:
          exact density calculated using eq.~(\ref{eq:tilde-n-def}).
          Other lines: density calculated by the random-wave-function algorithm
          with $N_{\rm R}=20$ (green dashed line)
          and $N_{\rm R}=1000$ (black dotted line) realizations.
        Dimensionless units defined
        in eqs.~(\ref{eq:dimensionless})--(\ref{eq:energy-unit}) are used.
        The sample size is equal to 30 dimensionless units, the
        discretization parameter is $a=0.1$, the temperature
        is $T=1$.\label{fig:algorithm1}}
\end{figure}

An example of the electron density distribution
in a one-dimensional white-noise potential is shown
in Fig.~\ref{fig:algorithm1}.
For convenience, we plot the reduced density $\tilde{n}(x,T)
= e^{-\beta\mu} n(x,T)$ that is independent of the chemical potential~$\mu(T)$.
Here, $x$ is the coordinate along the sample,
and the electrons are non-interacting.
The solid orange line depicts the exact function $\tilde{n}(x,T)$,
calculated from the exact eigenvectors $\varphi_i(x)$ and
eigenvalues $\epsilon_i$ of the single-particle Hamiltonian
via eq.~(\ref{eq:tilde-n-def}).
The dashed green line and the dotted black line are the reduced
electron densities calculated by our algorithm
using $N_{\rm R} = 20$ and $N_{\rm R} = 1000$ independent realizations,
respectively.
As seen from the figure, even with few iterations, $N_{\rm R}=20$,
the algorithm reasonably reproduces the shape of $\tilde{n}(x,T)$.
With $N_{\rm R} = 1000$, the electron density almost perfectly
reproduces the exact
result.

\begin{figure}[t]
	\includegraphics[width=\linewidth]{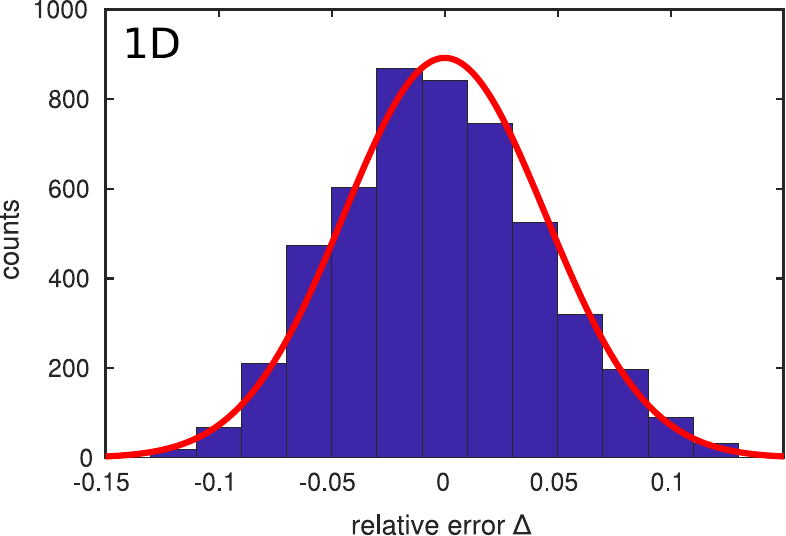}
	\caption{Distribution of the relative error $\Delta$
          of the random-wave-function algorithm with $N_{\rm R}=1000$
          in a sample with a one-dimensional
          white-noise potential. The quantity $\Delta$ is defined in
          eq.~(\ref{eq:Delta-def}).
          Blue bars: histogram of the numerically obtained
          distribution.
          Red line: Gaussian distribution with expectation value zero
          and variance $2/N_{\rm R}$.
          Parameters are the same as in Fig.~\ref{fig:algorithm1},
          except for the sample size that equals 500 dimensionless
          units.\label{fig:algorithm2}}
\end{figure}

In order to quantify the accuracy of the random-wave-function
algorithm with $N_{\rm R} = 1000$,
we calculate the distribution of the relative error $\Delta$,
as defined in eq.~(\ref{eq:Delta-def}). The histogram of $\Delta$
is represented in Fig.~\ref{fig:algorithm2}.
The red line in this figure shows a Gaussian distribution
with the mean-square value equal to $\sqrt{2/N_{\rm R}} \approx 0.045$,
according to eq.~(\ref{eq:Delta2-expected}).
The agreement between numerical (bars) and theoretical (line)
distributions demonstrates the correctness of the estimate~(\ref{eq:Delta2-expected})
for the accuracy of the random-wave-function algorithm.

\begin{figure}[t]
  \includegraphics[width=\linewidth]{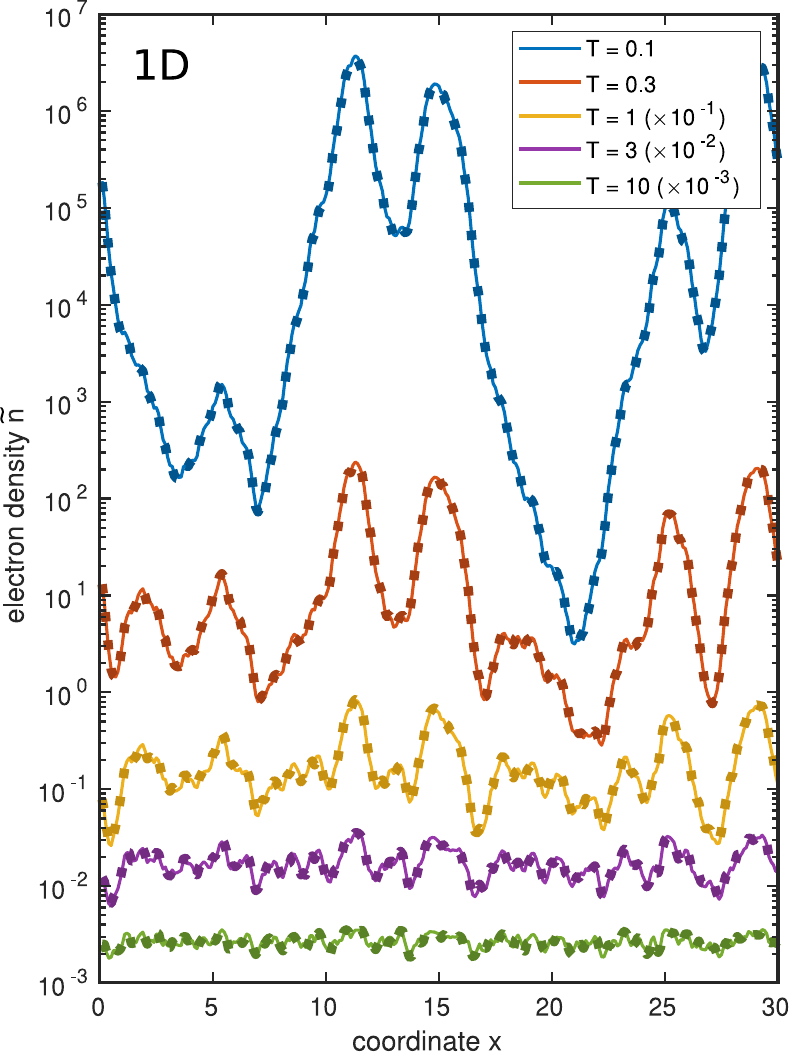}
	\caption{Comparison between exact (solid lines)
          and approximate (symbols) reduced electron density
          $\tilde{n}(x,T) = e^{-\beta\mu} n(x,T)$
          in a one-dimensional white-noise potential
          at different temperatures~$T$
          and $N_{\rm R} = 1000$ iterations.
          All other parameters are the same as in
          Fig.~\ref{fig:algorithm1}, except for the temperature.
          For clarity, the lowest three curves are scaled
          down by a multiplication by $10^{-3}$, $10^{-2}$ and $10^{-1}$,
          as indicated in the legend.\label{fig:algorithm3}}
\end{figure}

The only free parameter of the model is the dimensionless temperature $T$.
So far, we tested our algorithm for $T=T_0$ only.
Now, we apply it for a variety of temperatures, $0.1T_0 \le T \le 10T_0$.
The results for the reduced electron density $\tilde{n}(x,T)$
are shown in Fig.~\ref{fig:algorithm3}.
Lines represent the exact density calculated by a diagonalization
of the Hamiltonian and applying eq.~(\ref{eq:tilde-n-def}).
Dotted lines are obtained by the random-wave-function algorithm
with $N_{\rm R} = 1000$ independent realizations.
Just as in Fig.~\ref{fig:algorithm1}, we see that the algorithm enables a very
accurate calculation of the electron density.
The algorithm is thus seen to work for a broad range of temperatures.

\subsubsection{Two- and three-dimensional cases}

We perform similar numerical studies for 2D and 3D samples
of a random white-noise potential.
For the sake of comparison with the exact electron density,
we consider rather small samples,
with the number of grid nodes~$L$ less than 10.000.
This permits a complete numerical diagonalization of the Hamiltonian
to obtain the exact value of the electron density
via eqs.~(\ref{eq:n-via-tilde-n}) and~(\ref{eq:tilde-n-def})
for reference.
The size of the 2D sample is chosen to be $5 \times 5$ dimensionless units
($L = 2500$), and the size of the 3D sample is chosen to be
$2 \times 2 \times 2$ dimensionless units ($L = 8000$),
with grid period $a = 0.1$.

\begin{figure}[t]
	\includegraphics[width=\linewidth]{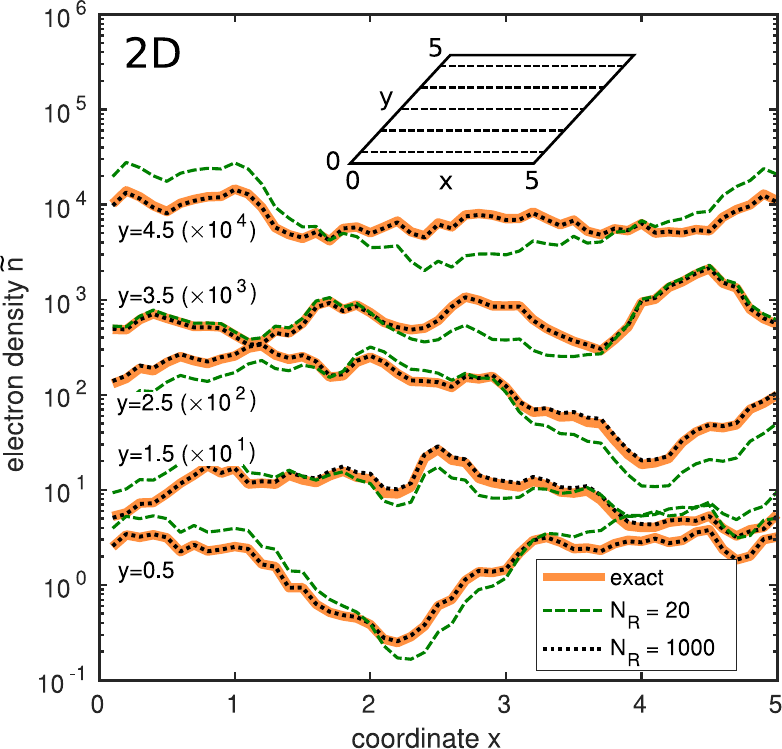}
	\caption{Reduced electron density $\tilde{n}(x,y,T)$
          in a two-dimensional sample with white-noise potential.
          We compare the exact reduced density (solid orange line)
          with the approximate
          reduced density obtained from the random-wave-function algorithm
          (dashed green line for $N_{\rm R} = 20$ and dotted black line for
          $N_{\rm R} = 1000$).
          Profiles along the $x$-axis with different values of
          coordinate~$y$ are shown (see inset). The sample size is $5\times5$
          dimensionless units, the discretization grid parameter is $a=0.1$,
          the temperature is $T=1$.
          Periodic boundary conditions apply.
          For clarity, profiles are multiplied by different coefficients,
          as indicated in the plot.\label{fig:algorithm4}}
\end{figure}

\begin{figure}[t]
	\includegraphics[width=\linewidth]{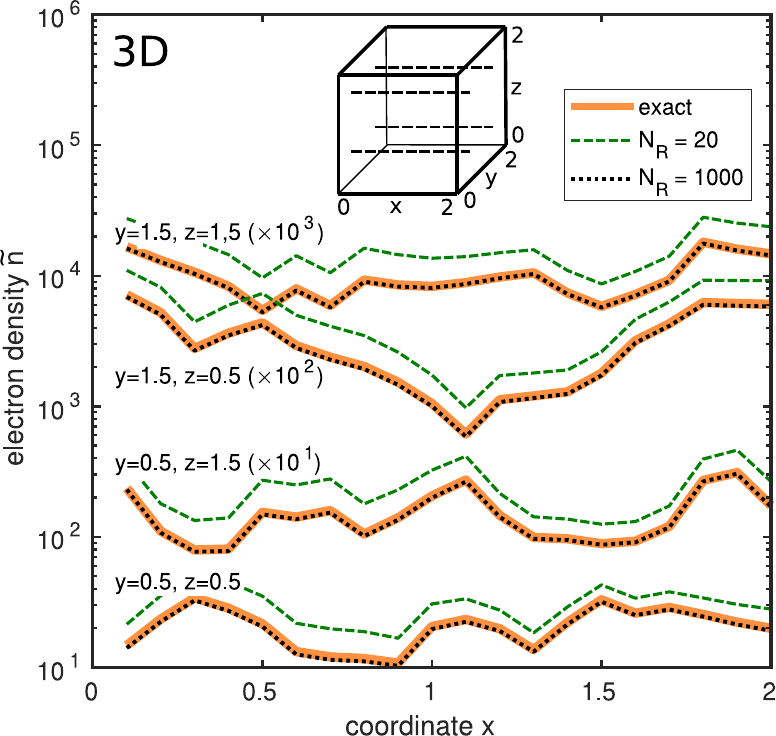}
	\caption{Reduced electron density $\tilde{n}(x,y,z,T)$
          in a three-dimensional sample with white-noise potential.
          Compared are
          the exact density (solid orange line)
          with that obtained by the random-wave-function algorithm
          (dashed green line for $N_{\rm R} = 20$ and dotted black line for
          $N_{\rm R} = 1000$).
          Profiles along the $x$-axis with different values of
          coordinates $y$ and $z$ are shown (see inset).
          Sample size is $2\times2\times2$ dimensionless units,
          the discretization grid parameter is $a=0.1$,
          the temperature is $T=1$.
          Periodic boundary conditions apply.
          For clarity, profiles are multiplied by
          different coefficients,
          as indicated in the plot.\label{fig:algorithm5}}
\end{figure}

For the 2D sample,
profiles of the reduced electron density $\tilde{n}(x,y,T)$
along the $x$-axis at several values of $y$ are shown in
Fig.~\ref{fig:algorithm4}.
Similarly, the $x$-profiles of the reduced density $\tilde{n}(x,y,z,T)$
in the 3D sample at fixed values of $y$ and $z$ are shown in
Fig.~\ref{fig:algorithm5}.
The color scheme in these figures is the same as in Fig.~\ref{fig:algorithm1}.
The exact density $\tilde{n}(\r,T)$ is plotted as a orange solid line,
and the densities obtained via the random-wave-function
algorithm with $N_{\rm R} = 20$
and $N_{\rm R} = 1000$ by a green dashed line and by a black dotted line,
respectively.

The question might arise on why the data in Fig.~\ref{fig:algorithm4} look qualitatively different to those in Fig.~\ref{fig:algorithm5}. While the dashed green line in Fig.~\ref{fig:algorithm4}  (2D case) appears to fluctuate around the exact results, it appears to be shifted upwards by a close to constant amount at each point in Fig.~\ref{fig:algorithm5} (3D case). This qualitative difference is due to the different sizes of the samples simulated in 2D case and in 3D case.

Figs.~\ref{fig:algorithm4} and~\ref{fig:algorithm5}
show that the random-wave-function algorithm provides qualitatively
correct results
already for a small number of iterations, $N_{\rm R} = 20$.
With $N_{\rm R}= 1000$ iterations, we can
reproduce the electron density very accurately.
Hence, our method to determine the density for a non-degenerate,
non-interacting electron gas is seen to work for 1D, 2D, and 3D
random potentials.
The RWF algorithm permits
to obtain the electron density in much larger samples
where a straightforward full diagonalization
of the Hamiltonian is no longer feasible.

\section{Effective potential}
\label{sec:eff-potential}

In this section we relate the electron density $n(\r,T)$ to a (quasi-)classical,
temperature-dependent effective potential $W(\r,T)$ for two reasons.
First, the random-wave-function algorithm requires
matrix multiplications with (sparse) $L\times L$ matrices where
$L$ is the number of grid points of the sample, and we are interested to find
a numerical method that scales more favorably in the system size.
Second, $W(\r,T)$ visualizes the potential landscape
in which classical particles would move so that concepts
like percolation theory can be applied.

\subsection{Quasi-classical effective potential}

We start from the well-known expression for the electron density $n(\r,T)$
of a non-degenerate electron gas, e.g.,
in the conduction band of a semiconductor,
\begin{equation}
  n(T) =
  N_c \exp\left( \frac{\mu - V}{k_{\rm B} T} \right)
  \quad \text{(constant potential)}\; ,
 \label{eq:n-constant-V}
\end{equation}
where $V$ is the energy of the conduction band edge, i.e.,
the electrons' potential energy, and $N_c$ is the effective density of states,
\begin{equation}
  N_c = 2 \left( \frac{m k_{\rm B} T}{2 \pi \hbar^2} \right)^{d/2}
  \label{eq:Nc-def}
\end{equation}
in $d$~dimensions. The relation~(\ref{eq:n-constant-V})
remains applicable
also in the case of a coordinate-dependent potential $V(\r)$
if the potential is smooth on the scale of the de-Broglie wavelength,
\begin{equation}
  n(\r,T) = N_c \exp\left[ \frac{\mu - V(\r)}{k_{\rm B} T} \right]
\quad \text{(smooth potential)}\; .
  \label{eq:n-smooth-V}
\end{equation}
In other words, electrons near the position $\r$ feel only the
local value $V(\r)$
of the potential. This approximation corresponds to the quasi-classical
description
of electron motion in a smooth potential $V(\r)$.
However, eq.~(\ref{eq:n-smooth-V}) cannot be applied when
the potential fluctuates significantly on length scales less or
equal to the de-Broglie wavelength. Since the particles are strongly
scattered by the
potential, the quasi-classical picture of a compact wave packet breaks down.

As a result of the quantum-mechanical treatment,
the electron density at some point $\r$ is affected by the potential not only at $\r$
but also in some extended vicinity. Therefore,
the electron density is a much smoother function of coordinates than the potential.
In this sense one might argue that the electron gas experiences
a `smoothed' potential. Therefore, instead of eq.~(\ref{eq:n-smooth-V}),
one might expect that the electron density obeys
\begin{equation}
  n(\r,T) = N_c \exp\left[ \frac{\mu - W(\r,T)}{k_{\rm B} T} \right]\;,
  \label{eq:n-via-W}
\end{equation}
where $W(\r,T)$ is the quasi-classical effective potential
in the case of a strongly fluctuating potential.
This potential is obtained from the actual potential $V(\r)$ by some appropriate
operation of `smoothing'.

We will consider eq.~(\ref{eq:n-via-W}) as the very \emph{definition\/}
of the quasi-classical effective potential $W(\r,T)$. Hence,
\begin{equation}
  W(\r,T) \overset{\text{def}}{=} \mu(T) - k_{\rm B} T \ln\frac{n(\r,T)}{N_c} \; .
  \label{eq:W-via-n}
\end{equation}
It is important to note that the
effective potential $W(\r,T)$ also depends on temperature.
This temperature dependence
is demonstrated numerically in Sect.~\ref{eq:white-noise-numerics}
for the example of a white-noise potential.
For this reason, the effective potential introduced here
differs from that of the Localization-Landscape Theory
where the effective potential $W_{\rm LLT}(\r)$ is independent of~$T$.

\subsection{Linear low-pass filter}
\label{sec:low-pass-filter}

To simplify the notation, in this subsection we drop
the temperature-dependence of all quantities.
Let us consider the
response $\delta n(\r)$ of the electron gas density
to a variation $\delta V(\r')$ of the potential. To first order in $\delta V$,
\begin{equation}
  \delta n(\r) = \int \d\r' \, \Gamma_n(\r,\r') \, \delta V(\r')
  + {\cal O}(\delta V^2) \;,
   \label{eq:delta-n-via-delta-V}
\end{equation}
where the linear-response function $\Gamma_n(\r,\r')$
can be defined as a functional derivative,
\begin{equation}
  \Gamma_n(\r,\r') = \frac{\delta n(\r)}{\delta V(\r')} \; .
   \label{eq:Gamma-n-def}
\end{equation}
Likewise, one can define a linear-response function $\Gamma_W$
for the effective potential $W$,
\begin{equation}
  \delta W(\r) = \int \d\r' \, \Gamma_W(\r,\r') \, \delta V(\r')
  + {\cal O}(\delta V^2) \;,
  \label{eq:delta-W-via-delta-V}
\end{equation}
or, equivalently, as a functional derivative,
\begin{equation}
  \Gamma_W(\r,\r') = \frac{\delta W(\r)}{\delta V(\r')} \; .
  \label{eq:Gamma-W-def}
\end{equation}
The relation between the response functions $\Gamma_n$ and $\Gamma_W$
can be readily found using the chain rule and eq.~(\ref{eq:W-via-n}),
\begin{equation}
  \Gamma_W(\r,\r') = \frac{\d W(\r)}{\d n(\r)} \; \frac{\delta n(\r)}{\delta V(\r')}
  = -\frac{k_{\rm B} T}{n(\r)} \, \Gamma_n(\r,\r') \;.
  \label{eq:Gamma-W-via-Gamma-n}
\end{equation}
In Section~\ref{universal-filter} we show how to evaluate these response
functions with the help of perturbation theory.

The function $\Gamma_W$ has a remarkable property: it is normalized to unity,
\begin{equation}  \label{eq:Gamma-W-normalization}
    \int \d\r' \, \Gamma_W(\r,\r') = 1\; .
\end{equation}
To see this, we insert a coordinate-independent variation
$\delta V(\r) = \text{const} = \delta V$
into eq.~(\ref{eq:delta-W-via-delta-V}) and obtain
\begin{equation}
  \delta W(\r) = \delta V \int \d\r' \, \Gamma_W(\r,\r')
  + {\cal O}(\delta V^2) \; .
  \label{eq:delta-W-const-delta-V}
\end{equation}
On the other hand, adding a constant $\delta V$ to the potential gives rise to the
same shift of all electron energy levels $\epsilon_i$. The electron density therefore
is changed by a factor of $\exp[-\delta V/(k_{\rm B} T)]$. According
to definition~(\ref{eq:W-via-n}), the effective potential $W(\r)$
acquires the constant shift
\begin{equation}
  \delta W(\r) = \delta V \; .
  \label{eq:delta-W-const-delta-V-2}
\end{equation}
The normalization condition~(\ref{eq:Gamma-W-normalization}) then follows from
the comparison of eqs.~(\ref{eq:delta-W-const-delta-V})
and~(\ref{eq:delta-W-const-delta-V-2}).

Since the function $\Gamma_W$ is properly normalized, it may play
the role of a filter function that produces the effective potential $W(\r)$
from the actual potential $V(\r)$ by the operation of convolution. However,
a filter function must depend on the difference $\r-\r'$ only, not on $\r$ and $\r'$
separately. For a random potential $V(\r)$  this is not the case because
the right-hand
side of eq.~(\ref{eq:Gamma-W-def}) depends on this potential
which breaks translational invariance.
Nevertheless, in a statistical sense we may argue
that the response function $\Gamma_W(\r,\r')$ that applies for a typical realization
of the disorder potential
can be approximated by some translationally invariant function,
\begin{equation}
  \Gamma_W(\r,\r') = \Gamma(\r-\r') \; ,
  \label{eq:Gamma-W-approx-Gamma}
\end{equation}
or, implying that the system also is isotropic in a statistical sense,
\begin{equation}
  \Gamma_W(\r,\r') = \Gamma(|\r-\r'|) \; .
  \label{eq:Gamma-W-approx-Gamma-iso}
\end{equation}
Due to translational invariance, the function $\Gamma$ does not depend on the
choice of the realization of the random potential. One can therefore integrate
the differential relation~(\ref{eq:delta-W-via-delta-V})
between two random realizations $V_1(\r)$ and $V_2(\r)$ which yields
$  U_1(\r)= U_2(\r)$ where
\begin{equation}
  U_i(\r)=  W_i(\r) - \int \d\r' \, \Gamma(|\r-\r'|) \, V_i(\r') \; ,
          \label{eq:W-integrated}
\end{equation}
that relates the effective potentials $W_1$ and $W_2$ to the actual potentials
$V_1$ and $V_2$, respectively. Consequently, $U_i(\r)\equiv C$,
whereby the constant $C$ does not depend on the choice of the realization.
In sum, the hypothesis~(\ref{eq:Gamma-W-approx-Gamma-iso})
gives rise to the following relation between the actual potential $V(\r)$
and the effective potential $W(\r)$,
\begin{equation}
  W(\r) = \int \d\r' \, \Gamma(|\r-\r'|) \, V(\r') + C \; .
  \label{eq:W-via-V}
\end{equation}
In words, \emph{the effective potential
  $W$ is a convolution of the real potential
  $V$ with the filter function $\Gamma$, plus some constant $C$}.
Note that the convolution with $\Gamma$ is nothing else but passing
the original potential through
the linear low-pass filter defined by $\Gamma$. For this reason,
the effective potential $W(\r)$ becomes much smoother
than the original potential $V(\r)$.
Since the effective potential depends also on the temperature, the filter function
$\Gamma$ and $C$ are temperature-dependent as well,
$\Gamma\equiv \Gamma(|\r-\r'|,T)$ and $C\equiv C(T)$.

In Section~\ref{eq:white-noise-numerics}, we test eq.~(\ref{eq:W-via-V})
numerically for the case of a white-noise random potential,
and ensure that it works surprisingly well, except for very low temperatures.
In Section~\ref{universal-filter}, we show that the
filter function $\Gamma$ can approximately be reduced to some universal
function $\Gamma_{\text{uni}}$ by the simple scaling relation
\begin{equation}
  \Gamma(r) \approx \lambdabar^{-d} \,
  \Gamma_{\text{uni}} (r/\lambdabar) \; ,
  \label{eq:Gamma-via-Gamma-uni}
\end{equation}
where $d$ is the spatial dimension, and $\lambdabar$
is the reduced de-Broglie wavelength of a charge carrier of mass $m$ and
kinetic energy $k_{\rm B} T$,
\begin{equation}
  \lambdabar = \frac{\hbar}{\sqrt{2m k_{\rm B} T}} \; ,
  \label{eq:lambda-de-Broglie}
\end{equation}
which is the thermal wave length up to a numerical factor.

\subsection{Numerical study of a white-noise potential}
\label{eq:white-noise-numerics}

In our numerical study of a white-noise potential,
we employ the model of white noise and the dimensionless system of units
described in Sect.~\ref{subsec:examples},
$\hbar = k_B = m = S = 1$.
The discretization parameter~$a$, the distance between grid nodes,
is chosen to be $a=0.1$. Periodic boundary conditions apply.
We consider a collection of 100 random samples of length 500 in 1D,
a collection of 20 samples of size $10\times10$ in 2D,
and a collection of 10 samples of size $10\times10\times10$ in 3D.
For each of these samples, we calculate the reduced electron density $\tilde{n}(\r,T)$.
In 1D and 2D, we use the definition of $\tilde{n}(\r,T)$, eq.~(\ref{eq:tilde-n-def}),
in which we employ the eigenfunctions $\varphi_i(\r)$ and eigenvalues $\epsilon_i$
from the numerical diagonalization of the Hamiltonian matrix.
In 3D, the Hamiltonian matrix is too large ($10^6 \times 10^6$)
for an exact numerical diagonalization so that we derive $\tilde{n}(\r,T)$
from the random-wave-function algorithm
described in Sect.~\ref{subsec:synthesis} where
we choose $N_{\rm R} \approx 1000 \sqrt{T}$ realizations so that
the accuracy for $\tilde{n}(\r,T)$ is about 5\% relative to the amplitude of spatial fluctuations.
The temperature-dependent effective potential $W(\r,T)$ is
then calculated from the reduced electron density as
\begin{equation}
  W(\r,T) = - k_{\rm B} T \ln\frac{\tilde n(\r,T)}{N_c} \; ,
  \label{eq:W-via-tilde-n}
\end{equation}
according to the definition~(\ref{eq:W-via-n}).
Since we have a collection of random potentials $V_{\rm R}(\r)$
and effective potentials $W_{\rm R}(\r,T)$,
we can test the validity of the filter function approach,
which is expressed mathematically by eq.~(\ref{eq:W-via-V}).

Before we apply the filter function approach,
some preliminary steps must be carried out.
First, we need to determine the value of the parameter $C$
in eq.~(\ref{eq:W-via-V}).
Second, it is necessary to estimate the characteristic range
of the filter function $\Gamma(r)$. Third,
we have to determine the optimal shape of this filter function.

\subsubsection{Constant parameter}

\begin{figure}[t]
	\includegraphics[width=\linewidth]{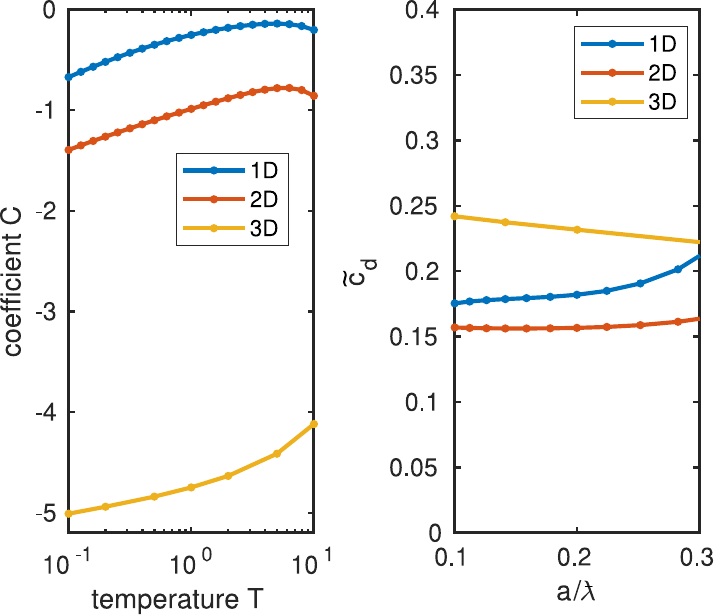}
	\caption{Left panel: parameter $C(T)$
          as function of the temperature $T$
          for systems with white-noise disorder in one, two, and three dimensions.
          The effective potential is defined by eq.~(\ref{eq:W-via-n}).
          The parameter $C$ is the average of the effective potential,
          $C = \langle W_{\rm R} \rangle_{\rm R}$ and used in eq.~(\ref{eq:W-via-V}).
          Dimensionless units ($\hbar = k_B = m = S = 1$)
          introduced in Section~\ref{subsec:examples} are used.
          Grid spacing $a=0.1$.
          Right panel: parameter function $\tilde{c}_d(a,T)$
          defined in eq.~(\ref{eq:Ctildeformula})
          as a function of $a/\lambdabar$. The limit $a/\lambdabar \to 0$
          defines the constants $c_d$.\label{fig:W_statistics}}
\end{figure}

To evaluate the parameter $C$, we note
that the mean value of the integral in eq.~(\ref{eq:W-via-V})
is equal to the mean value of $V$
because of the normalization property~(\ref{eq:Gamma-W-normalization})
of the filter function. Therefore,
when we take the mean values of both sides of eq.~(\ref{eq:W-via-V}), we obtain
\begin{equation}
  C = \langle W_{\rm R} \rangle_{\rm R} - \langle V_{\rm R} \rangle_{\rm R} \; ,
  \label{eq:C-value}
\end{equation}
where the brackets denote the average value over the collection of samples.
In the model used, the mean value of the potential $V_{\rm R}$ is equal to zero.
Hence, $C = \langle W_{\rm R} \rangle_{\rm R}$.
The resulting values of $C$ in the temperature range $0.1 \le T \le 10$
are plotted in Fig.~\ref{fig:W_statistics}, left part.
It is worth to note that the constant $C$ depends on the grid spacing~$a$.
The data in Fig.~\ref{fig:W_statistics} are calculated for $a=0.1$.

In order to extend the result to arbitrary values for~$a$,
we employ the following scaling argument that is corroborated
in appendix~\ref{app:B}.
The effective potential $W(\r,T)$ results from the random potential $V(\r)$
according to the linear mapping given in eq.~(\ref{eq:W-via-V}).
The contribution proportional to $V(\r)$ is obtained from first-order perturbation theory,
as explained in Sect.~\ref{universal-filter}.
Second-order perturbation contributes to $C$, after averaging over various
realizations of potential $V_{\rm R}(\r)$.
Consequently, $C$ is proportional to the potential squared and hence to
the parameter $S$ determined
in eq.~(\ref{eq:white-noise-def}).
The proportionality coefficient between~$C$ and~$S$ then follows from
the physical dimensions of these quantities.
While $C$ is energy, $S$ is energy squared times volume,
as evident from eq.~(\ref{eq:S-compositional}).
Therefore, $C_d(a,T)/S=  F_d(a/\lambdabar)/(k_{\textrm{B}}T \lambdabar^d)$.
where the thermal de-Broglie wavelength $\lambdabar$
is determined from eq.~(\ref{eq:lambda-de-Broglie}),
and the function $F_d(x)$ carries no units. It depends on the dimension~$d$
and on the ratio~$a/\lambdabar$.

As shown in appendix~\ref{app:B}, in the limit $x\to 0$, i.e., $a \ll \lambdabar$,
we have $F_d(x)= c_d f_d(x)$ with
$f_1(x)= -1$, $f_2(x)= \ln(0.30 x)$, and $f_3(x)= -1/x$.
Therefore,
\begin{equation}
  C_d(a,T)= S   \frac{c_d f_d(a/\lambdabar)}{k_{\textrm{B}}T \lambdabar^d} \; ,
\label{eq:CpropSformula}
\end{equation}
where $c_d$ are numerical factors.
To determine their values, we plot
\begin{equation}
\tilde{c}_d(a,T) =   \frac{C_d(a,T)k_{\textrm{B}}T \lambdabar^d}{S f_d(a/\lambdabar)}
\label{eq:Ctildeformula}
\end{equation}
in Fig.~\ref{fig:W_statistics}, right panel. The figure proves the scaling behavior
expressed in eq.~(\ref{eq:CpropSformula})
and permits to read off $c_1= 0.183$, $c_2= 1/(2\pi)\approx 0.159$, and $c_3= 0.255$
for the constants in $d= 1,2,3$ dimensions.

\begin{figure}[t]
	\includegraphics[width=\linewidth]{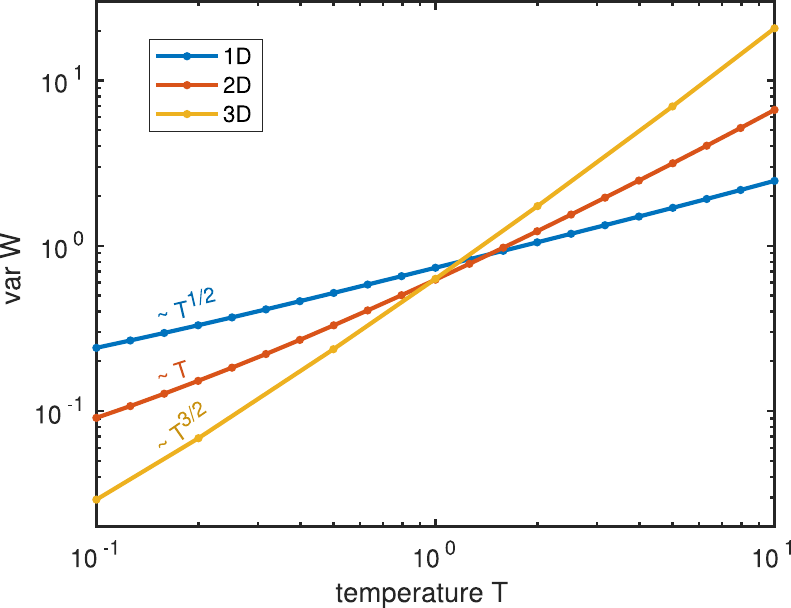}
	\caption{Variance of the effective potential $W_{\rm R}(T)$,
var~$W_{\rm R} = \langle W_{\rm R}^2 \rangle_{\rm R}
- \langle W_{\rm R} \rangle_{\rm R}^2$,
as function of the temperature~$T$ for systems with white-noise disorder
in one, two, and three dimensions.
          Dimensionless units ($\hbar = k_B = m = S = 1$)
          introduced in Section~\ref{subsec:examples} are used.
                    Grid spacing $a=0.1$.\label{fig:W_statisticsagain}}
\end{figure}

\subsubsection{Characteristic scale}

To determine the characteristic scale $\ell_\Gamma$
of the filter function $\Gamma$, we repeat the considerations from
Sect.~\ref{subsec:examples} and get the following estimate
similar to eq.~(\ref{eq:V-range}),
\begin{equation}
  \text{var}\,W_{\rm R} \simeq \frac{S}{\ell_\Gamma^d} \; ,
  \label{eq:l-estimate}
\end{equation}
where the symbol `var' denotes the variance, the square of the
standard deviation.
We show the variance of $W$ for different temperatures and spatial dimensions
in Fig.~\ref{fig:W_statisticsagain}.
Our numerical data clearly demonstrate that $\text{var}\,W \sim T^{d/2}$.
Inserting this dependence into eq.~(\ref{eq:l-estimate}), we see
that $\ell_\Gamma$ scales with temperature $T$
as $\ell_\Gamma \simeq 1/\sqrt T$.
This temperature dependence is the same as that of the thermal
wavelength $\lambdabar$, see eq.~(\ref{eq:lambda-de-Broglie}).
Specifically, $\lambdabar = 1/\sqrt{2T}$ in dimensionless units. Therefore,
\begin{equation}
  \ell_\Gamma \simeq \lambdabar \propto T^{-1/2}
  \label{eq:l-estimate-lambda}
\end{equation}
holds. The characteristic scale simply is the de-Broglie wave length.

\subsubsection{Shape of the low-pass filter}

To obtain the optimal shape of the low-pass filter $\Gamma(r)$,
we use the following variational scheme.
As a set of fitting parameters, we choose $\Gamma$ at $N\simeq10$
points $r_0 = 0, r_1, \ldots, r_N = 2/\sqrt T$.
The whole function $\Gamma(r)$ is restored from the values
$\Gamma(r_0), \ldots, \Gamma(r_N)$ using a cubic spline interpolation.
Two restrictions are imposed to the fitting parameters:
(i), $\Gamma(r_N)$ is fixed to zero and, (ii), the normalization
integral $\int \Gamma(|\r|) \, \d\r$ is fixed to unity.
The values $\Gamma(r_0), \ldots, \Gamma(r_{N-1})$ are varied
until the minimum of the relative error of the effective potential
\begin{equation}
  \mathfrak{E}[\Gamma]
  = \frac{\sqrt{\langle (W_\Gamma(\r) - W)^2 \rangle}}{\sqrt{\text{var}\,W}}
  \label{eq:relative-error-def}
\end{equation}
is reached. Here, $W$ is the effective potential calculated
from the electron density using eq.~(\ref{eq:W-via-tilde-n}),
and $W_\Gamma$ is calculated from the potential $V$ and the filter function
$\Gamma$ using eq.~(\ref{eq:W-via-V}). Here, the angular brackets imply
spatial integration.

\begin{figure}[t]
	\includegraphics[width=\linewidth]{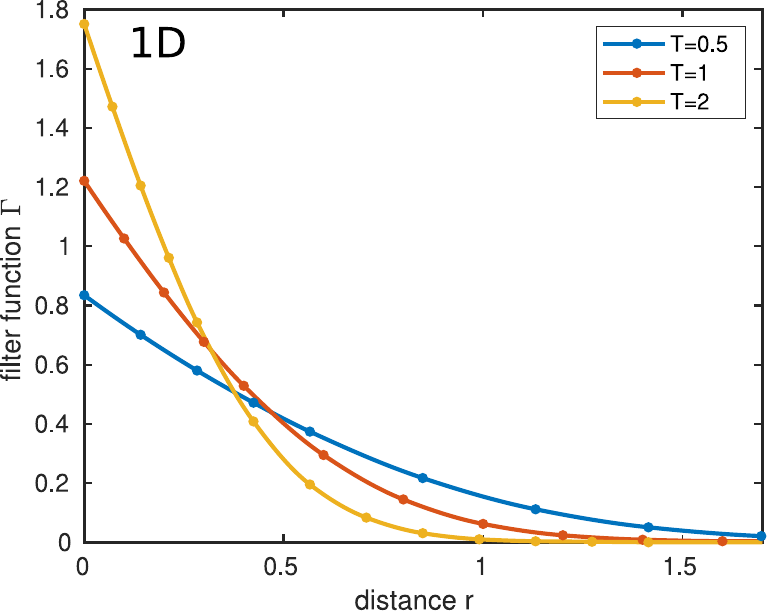}
	\caption{Filter functions $\Gamma(r)$ at different temperatures $T$
          in 1D systems with white-noise potential as obtained from the
          optimization procedure. The dots indicate values of $\Gamma(r)$
          that serve as fitting parameters.
          Dimensionless units ($\hbar = k_B = m = S = 1$)
          introduced in Sect.~\ref{subsec:examples}
          are used.\label{fig:1d_filter_functions}}
\end{figure}

\begin{figure}[t]
	\includegraphics[width=\linewidth]{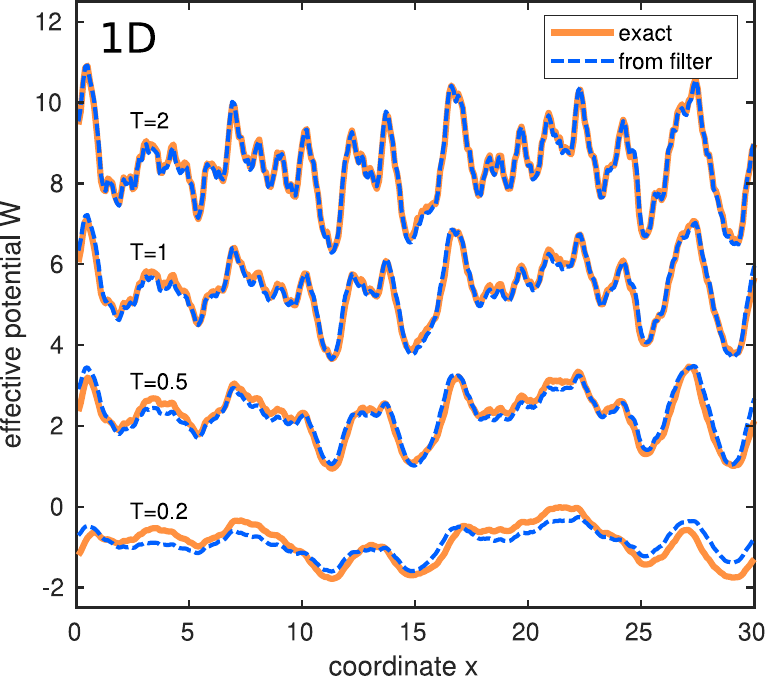}
	\caption{Comparison between the exact effective potential $W_{\rm R}(x)$
          (solid orange lines) and the filtered potential,
          see eq.~(\ref{eq:W-via-V}),
          (dashed blue lines) for a one-dimensional sample
          with while-noise potential.
          Filter functions at different temperatures $T$ are obtained
          by the optimization procedure. Dimensionless units
          ($\hbar = k_B = m = S = 1$) are used. The sample size is equal to 30
          dimensionless units, the discretization parameter is $a=0.1$.
          For clarity, curves for $T = 0.5$, $T = 1$, and $T = 2$
          are shifted upwards
          by 3, 6, and 9 units, respectively.\label{fig:1d_W}}
\end{figure}

Using this optimization procedure, we obtain the filter function $\Gamma(\r,T)$
in the temperature range $0.1 \le T \le 10$ for 1D, 2D, and 3D
white-noise potentials.
Examples of 1D filter functions for three different temperatures are shown
in Fig.~\ref{fig:1d_filter_functions}.
It is seen that the width of the filter function narrows
with temperature, in agreement with eq.~(\ref{eq:l-estimate-lambda}).
On the other hand, the shape of the filter function remains fairly unchanged
with varying the temperature. These properties of the filter function suggest
a universal filter function Ansatz, see eq.~(\ref{eq:Gamma-via-Gamma-uni}).

In Fig.~\ref{fig:1d_W},
we show an example of an effective potential $W_{\rm R}(x,T)$
in 1D at different temperatures (solid orange lines),
along with the estimates obtained by the filter-function approach,
eq.~(\ref{eq:W-via-V}) (dashed blue lines).
The agreement is quite good in the whole temperature range $T\gtrsim 0.2$.
At high enough temperatures, $T \ge 1$,
the filter-function approach reproduces the effective potential
almost perfectly.

\begin{figure}[t]
	\includegraphics[width=\linewidth]{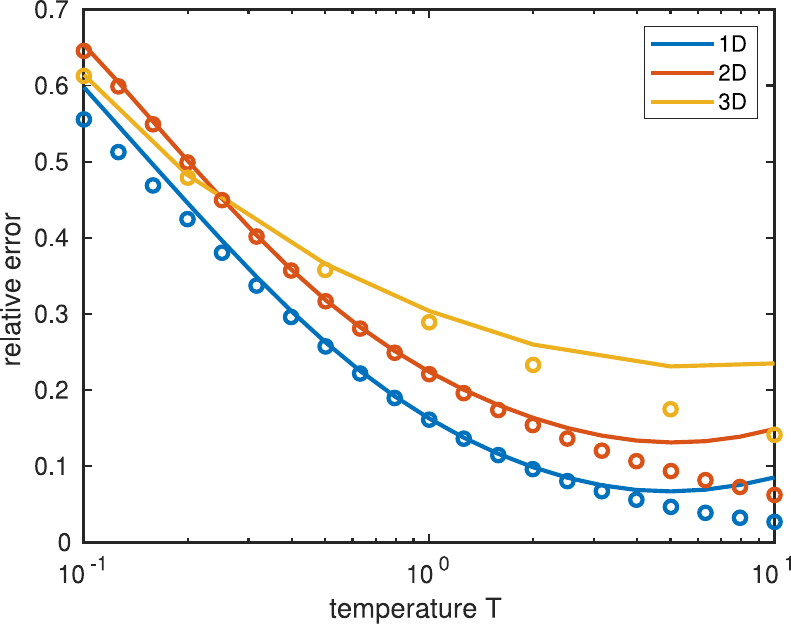}
	\caption{Relative error $\mathfrak{E}$ of the effective potential $W$
          obtained from convolution of the potential with the filter function,
          eq.~(\ref{eq:W-via-V}), at different temperatures $T$.
          Circles are related to the filter function obtained
          from the optimization procedure,
          lines correspond to the universal filter function
          considered in Sect.~\ref{universal-filter}.
          The values of relative error $\mathfrak{E}$ are obtained from
          eq.~(\ref{eq:relative-error-def}).
          Dimensionless units ($\hbar = k_B = m = S = 1$) are used.
          Discretization parameter $a=0.1$.\label{fig:filter_error}}
\end{figure}

To quantify the accuracy of our approach, we calculate
the relative error $\mathfrak{E}$ of the effective potential
obtained from by eq.~(\ref{eq:relative-error-def})
for different temperatures and spatial dimensions.
The results are shown in Fig.~\ref{fig:filter_error} by circles.
It is seen that the relative error decreases
with temperature.
In the low-temperature limit, the electron density is dominated by contributions
of rare low-energy eigenstates.
Therefore, the shape of the electron density distribution $n(\r,T)$
reproduces the probability distributions $|\varphi_i(\r)|^2$
of individual eigenstates~$\varphi_i(\r)$.
For this reason, the electron density distribution at very low temperatures
requires the solution of the Schr\"odinger equation,
and the filter-function approach becomes less reliable.
On the other hand, for high temperatures the role of the fluctuating potential
becomes negligible, the system under study acquires translational
and rotational symmetry. For high temperatures,
the validity of eq.~(\ref{eq:W-via-V}) for the filter-function
approach is guaranteed
because it follows from the universal relation~(\ref{eq:delta-W-via-delta-V}),
as explained in Sect.~\ref{sec:low-pass-filter}.

\subsection{Universal filter function}
\label{universal-filter}

As seen in Sect.~\ref{eq:white-noise-numerics}, the low-pass filter
representation
of the effective potential $W(\r,T)$ becomes more accurate
for higher temperatures.
Therefore, it is reasonable to search for a universal filter function
in the limit of infinitely
high temperature. This limit is characterized by neglecting
the random potential altogether.
Indeed, the high-temperature limit with a constant strength
of the random potential is equivalent to the limit of infinitely small strength
of the potential at constant temperature. For this reason, we develop
the universal low-pass filter (ULF) function in the case of zero potential where
the single-electron Hamiltonian $\hat{H}$ reduces to the
kinetic energy operator $\hat{T}$.

In this subsection, we first obtain an expression
for the response function $\Gamma_n(\r,\r')$ in terms of the matrix elements
of exponents of the Hamiltonian. This will be done in first-order
perturbation theory.
Then, we consider the case of zero potential, $\hat{H} = \hat{T}$,
and find analytical expressions for the response functions $\Gamma_n$
and $\Gamma_W$.
The latter plays the role of the filter function. Finally, we determine
an expression for the universal dimensionless
filter function $\Gamma_{\text{uni}}$. We numerically test
its accuracy at finite temperatures and disorder strengths.

\subsubsection{Linear response}

We start from expressions~(\ref{eq:n-via-tilde-n})
and~(\ref{eq:tilde-n-def}) that define
the electron density $n(\r,T)$ in a non-interacting,
non-degenerate electron gas.
They can be rewritten in operator notation as
\begin{equation}
  n(\r,T) = 2e^{\beta\mu} \langle \r | e^{-\beta\hat{H}} | \r \rangle \; ,
  \label{eq:n-via-exp-H}
\end{equation}
where $\hat{H}$ is the single-electron Hamiltonian, and the
position-space states $| \r \rangle$
are normalized such that
\begin{equation}
  \langle \r | \r' \rangle  = \delta(\r-\r') \; .
  \label{eq:r-norm}
\end{equation}
We are interested
in the variation $\delta n(\r)$ of the electron density in linear response
to a small variation $\delta V(\r)$ of the potential.
The potential variation can be
represented as a small addition $\widehat{\delta V}$ to the Hamiltonian,
\begin{equation}
  \widehat{\delta V} = \int \d\r' \delta V(\r') \,
  | \r' \rangle \langle \r' | \; .
  \label{eq:delta-V-operator}
\end{equation}
Therefore,
\begin{equation}
  \delta n(\r,T) = 2e^{\beta\mu} \langle \r |
  \left[ e^{-\beta(\hat{H}+\widehat{\delta V})} - e^{-\beta\hat{H}} \right]
  | \r \rangle  \; .
  \label{eq:delta-n-via-exp-H}
\end{equation}
The difference in square brackets can be evaluated via first-order
perturbation theory. Neglecting higher-order infinitesimals we find
\begin{equation}
  e^{-\beta(\hat H+\widehat{\delta V})} - e^{-\beta\hat H}
  = - \int_0^\beta \d\xi \, e^{-\xi\hat H} \,
  \widehat{\delta V} \, e^{-(\beta-\xi)\hat H} \; .
  \label{eq:perturb}
\end{equation}
Inserting eqs.~(\ref{eq:delta-V-operator}) and~(\ref{eq:perturb})
into eq.~(\ref{eq:delta-n-via-exp-H}), we obtain
\begin{eqnarray}
	\delta n(\r,T) &=& -2e^{\beta\mu} \int\limits_0^\beta \d\xi \int \d\r'
	\langle \r | e^{-\xi\hat H} | \r' \rangle
        \label{eq:delta-n-perturb} \\[-6pt]
        && \hphantom{-2e^{\beta\mu} \int\limits_0^\beta \d\xi \int \d\r' }
\times        \delta V(\r') \langle \r' |
        e^{-(\beta-\xi)\hat H} | \r \rangle \; .
     \nonumber
\end{eqnarray}
A comparison between the eq.~(\ref{eq:delta-n-perturb})
and eq.~(\ref{eq:delta-n-via-delta-V}) provides
the following expression for the linear response function $\Gamma_n(\r,\r',T)$,
\begin{equation}
\Gamma_n(\r,\r',T) = -2e^{\beta\mu} \int\limits_0^\beta \d\xi
\langle \r | e^{-\xi\hat H} | \r' \rangle \langle \r' | e^{-(\beta-\xi)\hat H}
| \r \rangle
\label{eq:Gamma-n-perturb}
\end{equation}
within first-order perturbation theory.

\subsubsection{Linear response for free electrons}

Now we turn to the particular case of free electrons, where the Hamiltonian
reduces to the kinetic energy operator $\hat{T}$,
\begin{equation}
  \hat{H} = \hat{T} = \sum_{\k} \frac{\hbar^2|\k|^2}{2m}
  | \k \rangle  \langle \k | \;,
  \label{eq:H-only-kinetic}
\end{equation}
where the wave-number states $| \k \rangle$ are related to
the position-space states~$| \r \rangle$ as
\begin{equation}
  \langle \r | \k \rangle = \frac{e^{{\rm i}\k\cdot\r}}{\sqrt{\Omega}} \; .
  \label{eq:k-r}
\end{equation}
We insert eqs.~(\ref{eq:H-only-kinetic}) and~(\ref{eq:k-r})
into eq.~(\ref{eq:Gamma-n-perturb}) to find
\begin{eqnarray}
  \Gamma_n(\r,\r',T)
  &=& -\frac{2e^{\beta\mu}}{\Omega^2} \sum_{\k,\k'} \int\limits_0^\beta \d\xi
F(\xi,\k,\k') \nonumber \\
F(\xi,\k,\k') &=&
\exp\left[   \frac{\hbar^2(\xi-\beta)|\k'|^2 - \xi|\k|^2}{2m}\right]\!
    e^{{\rm i}(\k-\k')\cdot(\r-\r')}\, .\nonumber \\
  \label{eq:Gamma-n-perturb-only-kinetic-1}
\end{eqnarray}
We perform the integral over $\xi$, replace the sums over $\k$ by integrations
over $\d\k \times \Omega/(2\pi)^d$, and discard the imaginary part to arrive
at an analytical expression for the response function,
\begin{eqnarray}
  \Gamma_n(\r,\r',T) &=&
  \frac{4me^{\beta\mu}}{\hbar^2(2\pi)^{2d}} \iint \d\k \, \d\k' G(\k,\k')\\
  G(\k,\k') &=&
  \left[\exp\left(-\frac{\beta\hbar^2|\k|^2}{2m}\right)-
    \exp\left(-\frac{\beta\hbar^2|\k'|^2}{2m}\right)\right]\nonumber \\
  && \times  \frac{\cos \left[ (\k-\k')\cdot(\r-\r') \right]}{|\k|^2 - |\k'|^2}
  \; ,
  \label{eq:Gamma-n-perturb-only-kinetic-2}
\end{eqnarray}
where the integrand at $|\k| = |\k'|$ is to be resolved by L'H\^ospital's rule.

The response function $\Gamma_W(\r,\r',T)$ for the effective potential
can be calculated
from eq.~(\ref{eq:Gamma-W-via-Gamma-n}).
In the absence of an external potential,
we use eq.~(\ref{eq:n-constant-V}) for the electron density $n(T)$
in a constant potential $V=0$. As a result, we obtain
the function
\begin{eqnarray}
	\Gamma_W(\r,\r')
	&=& \frac{\lambdabar^d}{2^d \pi^{3d/2}} \iint \d\k \,\d\k'
 \frac{e^{-|\lambdabar\k|^2}-e^{-|\lambdabar\k'|^2}}{|\lambdabar\k'|^2 -|\lambdabar\k|^2}
        \nonumber \\
      &&  \hphantom{\frac{\lambdabar^d}{2^d \pi^{3d/2}} \iint}
        \times \cos \left[ (\k-\k')\cdot(\r-\r') \right]  \; ,\nonumber \\
\label{eq:Gamma-perturb}
\end{eqnarray}
where we employ de-Broglie wavelength $\lambdabar$
from eq.~(\ref{eq:lambda-de-Broglie}).

Eq.~(\ref{eq:Gamma-perturb}) provides the filter function
$\Gamma(|\r-\r'|) \equiv \Gamma_W(\r,\r')$, introduced
in eq.~(\ref{eq:Gamma-W-approx-Gamma-iso}).
After substitutions $\lambdabar\k = \q$, $\lambdabar\k' = \q'$,
we arrive at eq.~(\ref{eq:Gamma-via-Gamma-uni}) that expresses
the temperature-dependent filter function $\Gamma$ via a universal,
dimensionless function $\Gamma_{\text{uni}}$.
This universal function, obtained with the help
of eq.~(\ref{eq:Gamma-perturb}),
is equal to
\begin{eqnarray}
	\Gamma_{\text{uni}}(|\tilde\r|) &=&
	\frac{1}{2^d \pi^{3d/2}} \iint \d\q \, \d\q'
	\left(e^{-|\q|^2} - e^{-|\q'|^2}\right)\nonumber \\
        &&\hphantom{	\frac{1}{2^d \pi^{3d/2}} \iint}
	\times \frac{\cos \left[ (\q-\q')\cdot\tilde\r \right]}{
          |\q'|^2 - |\q|^2 }   \; ,
        \label{eq:Gamma-uni-perturb}
\end{eqnarray}
where $\tilde\r = \r/\lambdabar$ is a dimensionless radius vector.

\begin{figure}[t]
	\includegraphics[width=\linewidth]{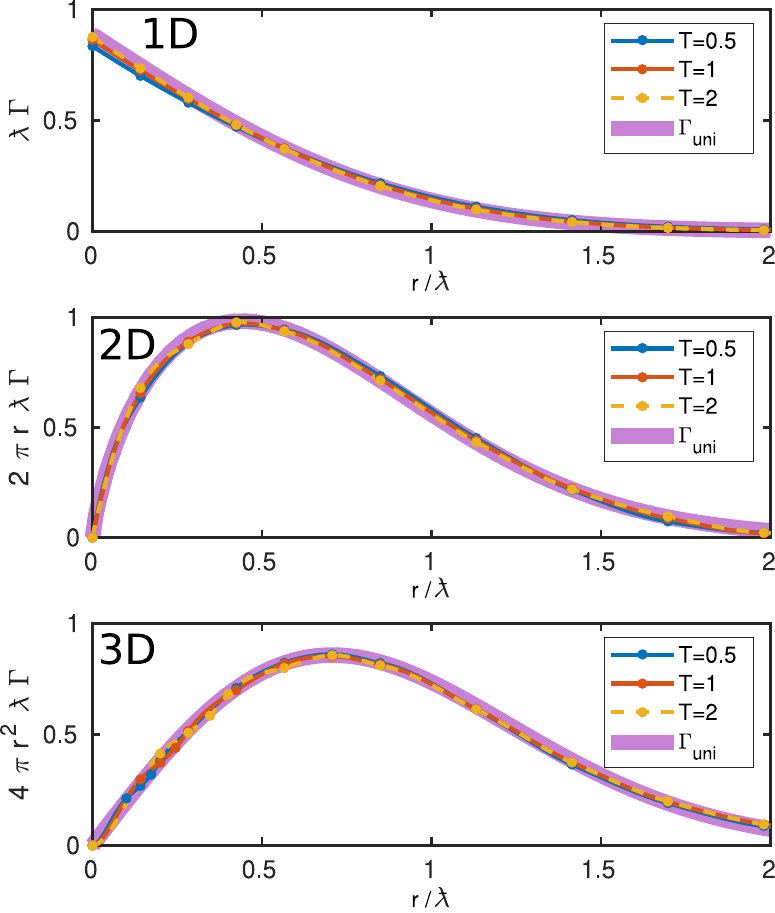}
	\caption{Radial filter functions $\Gamma_W^{(d)}(r/\lambdabar)$ ($d = 1,2,3$
         from top to bottom),
          obtained from
          the optimization procedure in Sect.~\ref{eq:white-noise-numerics},
          plotted versus $r/\lambdabar$,
          in comparison with the universal filter function $\Gamma_{\text{uni}}$
          (thick violet line).
          Dimensionless units ($\hbar = k_B = m = S = 1$) are used.
          $\lambdabar = 1/\sqrt{2T}$ is the reduced de-Broglie wavelength
          at kinetic energy equal to $T$. The discretization parameter is
          $a=0.1$.\label{fig:universal_filter}}
\end{figure}

This expression can be simplified by means of Fourier transform.
The Fourier image $\hat\Gamma_{\text{uni}}(k)$ of the universal filter function
is defined as
\begin{equation}
  \hat\Gamma_{\text{uni}}(|\k|) = \int \d\r \, \Gamma_{\text{uni}}(|\r|) \,
  e^{{\rm i}\k\cdot\r} \; .
  \label{eq:hat-Gamma-uni-def}
\end{equation}
The substitution of eq.~(\ref{eq:Gamma-uni-perturb}) into
eq.~(\ref{eq:hat-Gamma-uni-def}) gives rise to the simple result
\begin{equation}
  \hat\Gamma_{\text{uni}}(k) = \frac{\sqrt\pi}{k} \, e^{-k^2/4} \, \text{erfi}(k/2) \; ,
  \label{eq:hat-Gamma-uni}
\end{equation}
where erfi is the imaginary error function,
\begin{equation}
  \text{erfi}(\xi) = -{\rm i} \, \text{erf}({\rm i}\xi)
  = \frac{2}{\sqrt\pi} \int_0^\xi e^{t^2} \d t \; .
  \label{eq:erfi-def}
\end{equation}
Expression (\ref{eq:hat-Gamma-uni}) resembles the filter function, which has been used by Steinerberger~\cite{Steinerberger2021} in order to construct the effective potential from the initial random one. Note that the function $\hat\Gamma_{\text{uni}}(k)$
does not depend on the spatial dimension.
The Fourier image $\hat\Gamma(k)$ of the temperature-dependent filter function
$\Gamma(r)$ is obtained from eq.~(\ref{eq:Gamma-via-Gamma-uni}),
\begin{equation}
  \hat\Gamma(k) \approx \hat\Gamma_{\text{uni}}(\lambdabar k) \; .
  \label{eq:hat-Gamma-via-hat-Gamma-uni}
\end{equation}

In Fig.~\ref{fig:universal_filter}, we compare the filter functions $\Gamma(r)$
obtained by the optimization procedure in Sect.~\ref{eq:white-noise-numerics}
with the universal function $\Gamma_{\text{uni}}(r)$.
For this purpose, we multiply the temperature-dependent filter functions
by $\lambdabar^d$, and plot them versus $r/\lambdabar$.
The universal function, obtained by inverse Fourier transformation
of~(\ref{eq:hat-Gamma-uni}),
is shown by thick violet lines.
A dimensionless system of units is used,
in which $\hbar = k_B = m = S = 1$ and hence $\lambdabar=1/\sqrt{2T}$.
It is seen that in all dimension
the scaled filter functions approximately follow the universal function
$\Gamma_{\text{uni}}$. Therefore, the scaling
property~(\ref{eq:Gamma-via-Gamma-uni})
of the filter functions is numerically shown,
and the expression~(\ref{eq:hat-Gamma-uni}) for the universal filter function
is validated.

\subsubsection{Effective potentials using a universal low-pass filter}
\label{sec:low_pass_filter}

Using a universal low-pass filter (ULF) provides a practical numerical method
for the
calculation of the electron density distribution $n(\r,T)$
for a given realization of the white-noise potential $V_{\rm R}(\r)$.
The method consists of the following simple steps.
\begin{itemize}
\item[(1)]
  Calculate the Fourier image $\hat{V}_{\rm R}(\k)$
  of the random potential by means of a fast-Fourier-transform (FFT) algorithm.
\item[(2)] Multiply the function $\hat{V}_{\rm R}(\k)$ by $\hat{\Gamma}(|\k|)$,
  where $\hat\Gamma(|\k|)$ is given by eqs.~(\ref{eq:lambda-de-Broglie}),
  (\ref{eq:hat-Gamma-uni}), and~(\ref{eq:hat-Gamma-via-hat-Gamma-uni}).
\item[(3)] Perform the inverse Fourier transform of the product
  $\hat{V}(\k) \hat{\Gamma}(|\k|)$ by FFT, and add the constant $C(T)$
   to the result,  see eq.~(\ref{eq:CpropSformula}).
  The output of the third step is the effective potential $W_{\rm R}(\r,T)$
  which follows from eq.~(\ref{eq:W-via-V}) and from the fact that
  inverse Fourier transform converts a product into a convolution.
\item[(4)] Calculate the electron density $n(\r,T)$ from the effective
  potential $W_{\rm R}(\r,T)$
  using eq.~(\ref{eq:n-via-W}).
\end{itemize}
Since the numerical effort in FFT scales proportional to $L\ln(L)$,
the effective potential
$W(\r,T)$ and the particle density $n(\r,T)$ can be calculated for
mesoscopically large systems from a microscopic random potential $V(\r)$.

\begin{figure}[t]
	\includegraphics[width=\linewidth]{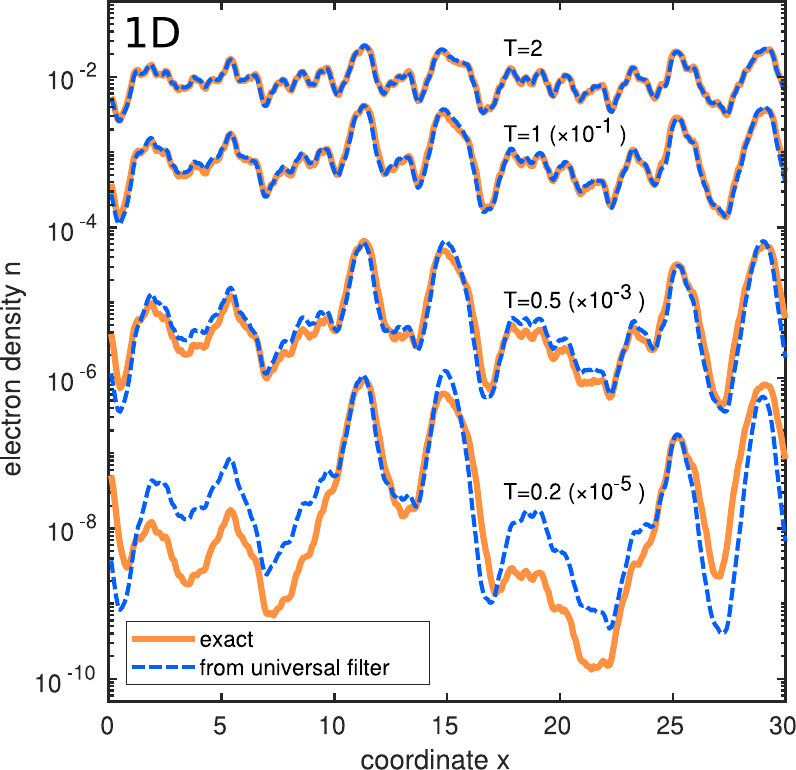}
	\caption{Electron density $n(x,T)$ in a one-dimensional white-noise potential
          at different temperatures $T$: exact (solid orange lines) and
          universal low-pass filter (ULF) (dashed blue lines).
          The parameters are the same as
          in Fig.~\ref{fig:algorithm1}, except for the temperature $T$.
          For clarity, the curves related to $T=0.2$, $T=1$ and $T=2$
          are multiplied by $10^{-4}$, $10^2$ and $10^3$, respectively,
          as indicated in the plot; particle concentration $n = 0.01$.\label{fig:1d_n}}
\end{figure}

\subsubsection{Numerical tests}
\label{subsec:univ_filter_Numerical tests}

Lastly, we present numerical tests of our method.
Fig.~\ref{fig:filter_error} shows the relative error
of the calculated effective potential $W(\r,T)$
with the global filter (lines) in comparison with the same calculation using
filter functions
obtained numerically by an optimization procedure (circles).
It is seen that for a broad range of temperatures,
the universal filter function provides as accurate results as using
the numerically
optimized one.
An exception is the case of very high temperatures,
presumably due to discretization errors.

\begin{figure}[b]
  \includegraphics[width=\linewidth]{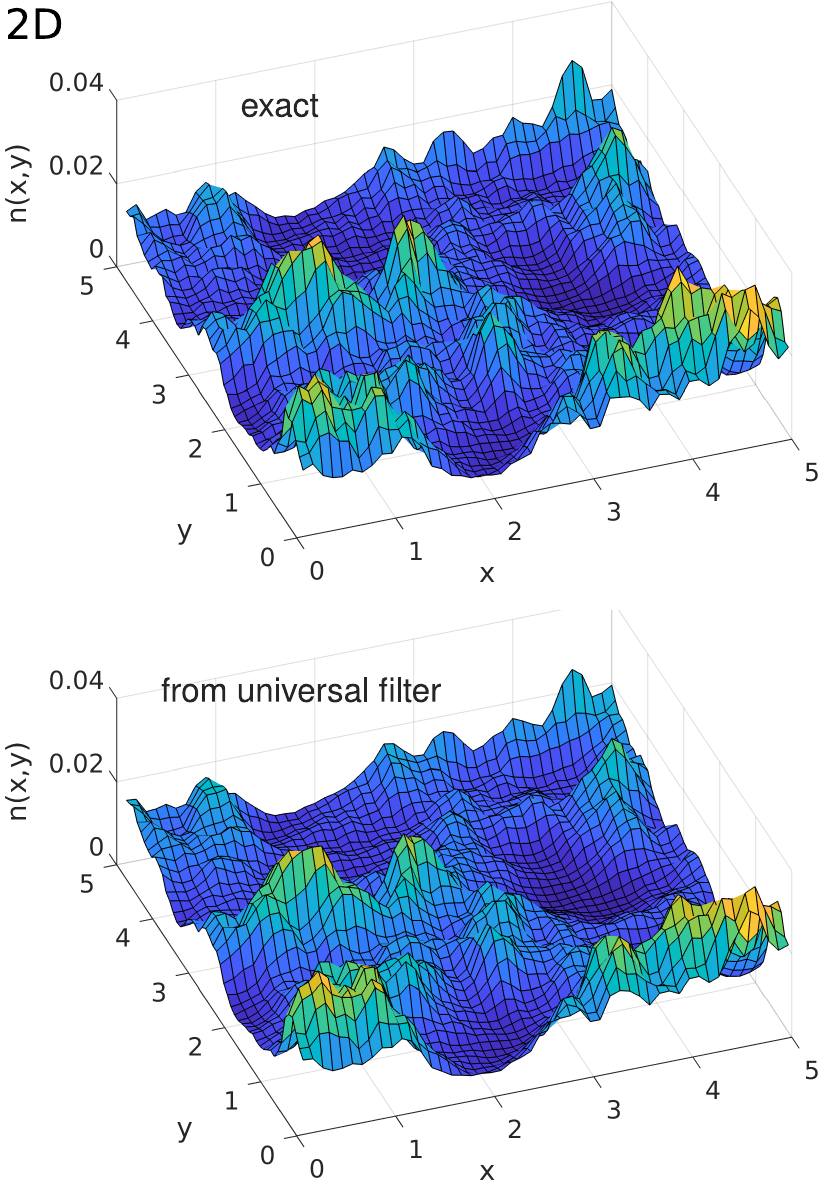}
  \vspace*{3pt}
	\caption{Comparison between the exact electron
          density $n(x,y,T)$ (upper part) and that obtained
          by using the universal filter function (lower part)
          in a two-dimensional white-noise potential.
          The sample size is $5\times5$ dimensionless units,
          the discretization parameter is $a=0.1$, the temperature is $T=1$,
          the particle concentration is $n=0.01$.
          Periodic boundary conditions apply.\label{fig:2d_n}}
\end{figure}

We compare the electron density distributions $n(\r,T)$
obtained by different numerical techniques.
Expressions~(\ref{eq:n-via-tilde-n}),
(\ref{eq:n-from-algorithm}), and (\ref{eq:n-via-W})
determine $n(\r,T)$ up to the factor $e^{\beta\mu}$,
where $\mu(T)$ is the chemical potential.
To remedy this ambiguity, we fix
the average electron concentration $n=N/\Omega$
at $n=0.01$.
Here, $N$ is the number of electrons, and $\Omega$ is the sample volume.
This concentration is much less than $N_c$ which permits to treat
the electron gas as non-degenerate.

Fig.~\ref{fig:1d_n} clearly demonstrates in 1D that the method
is fairly accurate
at temperatures $T \ge 0.5$. Here, the temperature is expressed
in dimensionless units
introduced in Sect.~\ref{subsec:examples}. Even for much smaller temperatures,
where the electron density $n(x,T)$ fluctuates by several orders of magnitude,
the universal-filter approach correctly predicts the positions
of maxima and minima of $n(x,T)$.

\begin{figure}[t]
  \includegraphics[width=\linewidth]{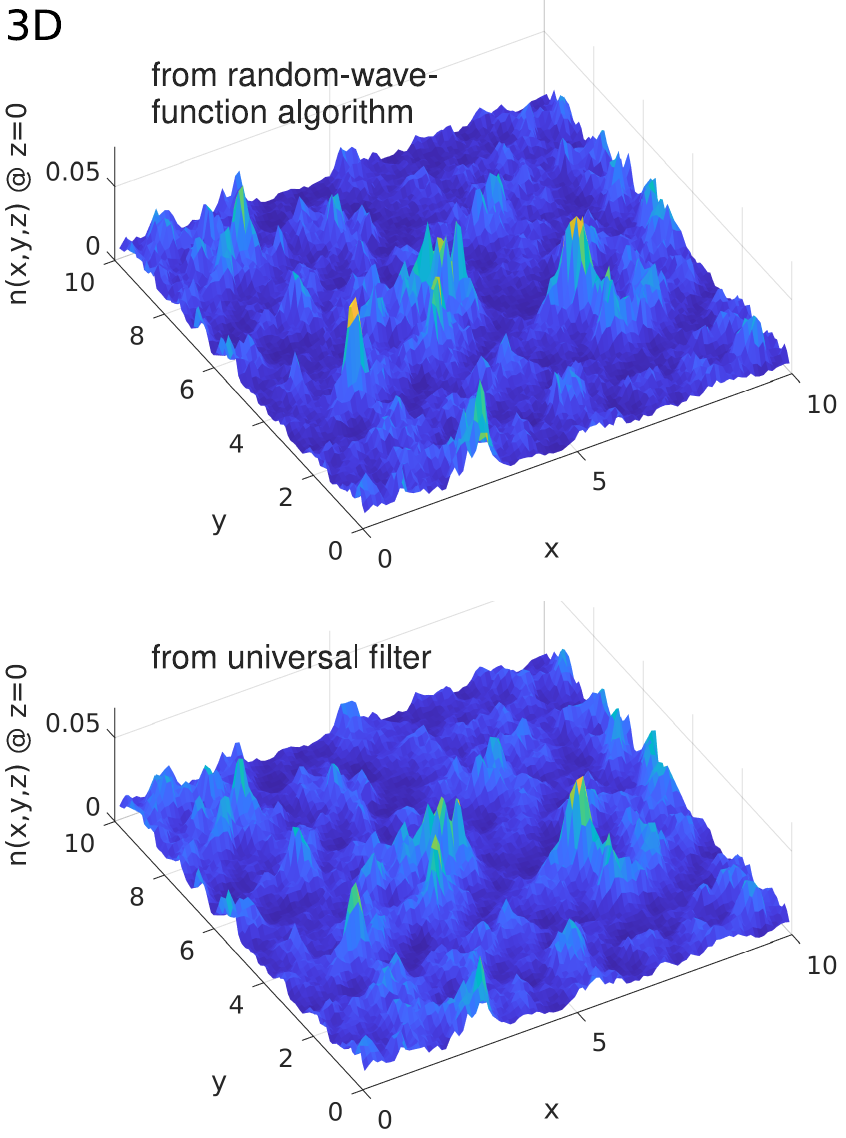}
	\caption{Comparison between the exact electron
          density $n(x,y,z=0,T)$ (upper part) and that obtained
          by using the universal filter function (lower part)
          in a three-dimensional white-noise potential.
          The sample size is $10\times10\times10$ dimensionless units,
          the discretization parameter is $a=0.1$, the temperature is $T=2$,
                    the particle concentration is $n=0.01$.
          Periodic boundary conditions apply.\label{fig:3d_n}}
\end{figure}

Fig.~\ref{fig:2d_n}
shows an application for a 2D system
with a white-noise potential at $T=1$.
 Since we fix the concentration $n=0.01$ and
work with a single particle species,
the effect of the constant $C(T)$ can be absorbed in the chemical potential,
$\tilde{\mu}(T)\equiv \mu(T)-C(T)$.
As can be seen from Fig.~\ref{fig:2d_n},
the exact result (upper plot) is accurately reproduced
by the calculation based on the universal filter function (lower plot).

Figure~\ref{fig:3d_n} shows an application for a 3D system
with a white-noise potential at $T=2$. The cross section of $n(x,y,z,T)$
at $z=0$ is plotted. The upper plot depicts the electron density distribution
$n(x,y,z,T)$ obtained by the random-wave-function algorithm with
relative accuracy 5\%.
The lower plot depicts the result obtained by the universal filter function.
The agreement between the two plots is good.

At first glance, these results are quite surprising because
the ULF method is based on perturbation theory.
In particular, the expression~(\ref{eq:Gamma-uni-perturb})
for the universal filter function $\Gamma_{\text{uni}}(r)$ is obtained
for high temperatures or small potential fluctuations.
One can expect that the perturbative approach must be restricted
to parameter sets that correspond to small density fluctuations.
However, it is seen that the approach works well even when
the density varies by more than an order of magnitude.

We surmise that the surprising success of the effective potential
in describing the local particle density at fairly low temperatures, $T\gtrsim 0.2 T_0$,
is related to the fact that $n(\r,T)$ depends on $W(\r,T)$ exponentially.
As in statistical physics, the error
appears to be proportional to higher-order cumulants in the expansion in $V(\r)/T$.
Cumulants often decay faster with temperature than the corresponding
coefficients of the bare series expansion because $n$th-order cumulants
describe true $n$-point correlations that cannot be factorized into
smaller subclusters.

\section{Comparison with Localization-Landscape Theory (LLT)}
\label{sec:ComparisonwithLLT}

In this section, we compare our results to those obtained from the
Localization-Landscape Theory (LLT) for the charge carrier concentration $n(\r,T)$.
In addition, we discuss the consequences of the different
approximate techniques of calculating $n(\r,T)$
for charge transport in disordered media at elevated temperatures.

The aim of the potential filtering with the universal low-pass filter (ULF) developed
in Sect.~\ref{sec:eff-potential} and that of the random-wave-function (RWF)
approach developed in Sect.~\ref{sec:density-calc} is to speed up calculations
compared to solving the Schr\"odinger equation.
The same aim has been targeted in the recently developed
LLT~\cite{Arnold2016,LL1_2017,LL2_2017,LL3_2017,WeisbuchNakamura2021}.

\subsection{Localization-Landscape Theory}

In the LLT, the right-hand side of the Schr\"odinger equation
is replaced with a constant to arrive at the landscape equation
\begin{equation}
 \hat{H}u(\r)=\left(-\frac{\hbar^2}{2m}\nabla^2+V(\r)\right)u(\r)=1 \; ,
  \label{eq:LLT}
\end{equation}
where $V(\r)$ is the disorder potential and $u(\r)$ is
localization landscape for appropriate boundary conditions.
The inverse of the landscape, $W(\r) \equiv 1/u(\r)$,
can be interpreted as a semi-classical effective confining potential
that determines the strength of the confinement
as well as the long-range decay of the quantum states.

\subsubsection{Positivity condition}

An important condition for the applicability of the LLT is that
the Hamiltonian $\hat{H}$ is a positive operator~\cite{LL1_2017}.
When solving the Schr\"odinger equation~(\ref{eq:Schroedinger}),
the potential $V(\r)$
experienced by the quantum particle can be defined up to a
constant~$K$. If one shifts the potential by $K$, then the
resulting eigenenergies~$\epsilon_i$ are shifted by the same constant~$K$.
However, this invariance does not hold for the landscape $u(\r)$. If
$u(\r)$ is the solution to eq.~(\ref{eq:LLT}),
then the solution $u_K(\r)$ corresponding to the same potential shifted by
the constant~$K$ satisfies~\cite{LL1_2017}
\begin{equation}
 -\frac{\hbar^2}{2m} \Delta u_K(\r) + \left[V(\r) + K\right]u_K(\r)=1 \; ,
  \label{eq:LLT_K}
\end{equation}
and $u(\r)\neq u_K(\r)$ for general $K\neq 0$.
In the LLT, the constant $K$ is chosen as small as possible
in such a way that the Hamiltonian
remains a positive operator~\cite{LL1_2017}, and the
electron density $n(\r,T)$ should be obtained from the effective potential
$W_K(\r) \equiv 1/u_K(\r)$ via eq.~(\ref{eq:n-via-W}).

Below we discuss that the choice of~$K$ drastically affects
the predictions of the LLT approach. In the calculations below, we choose
$K = -1.05 V_{\rm min}$, where $V_{\rm min}$ is the absolute minimum of $V(\r)$.

\subsubsection{Computational effort}

The exact, complete solution of the Schr\"odinger equation is computationally
demanding for large systems, and approximate techniques are mandatory
for random media in two and three dimensions.
The RWF requires the repeated application of the Hamiltonian onto a random
wave function. The time consumption of the latter approach depends on the system
size and on the number of realization used for the averaging procedure.
Therefore, the accuracy of the RWF approach can be systematically improved
by increasing the number of stored wave functions
and realizations of the impurity potentials for the price of increasing
computational resources. Thus, mesoscopically large systems cannot be
treated using the RWF.

The LLT is based on a solution of a system of linear equations.
Compared to solving the Schr\"odinger equation
self-consistently,
the LLT speeds up the calculations by two to three orders
of magnitude~\cite{Yun_Urbach2022}.
However, the ULF approach requires only Fast Fourier Transformation, which
does not consume noticeable computational time.
Below we compare the accuracy of all
three approximate approaches
in $d=1$ and $d=2$ and show that the ULF is superior to the LLT
not only in speed but also in precision.

\begin{figure}[b]
	\includegraphics[width=\linewidth]{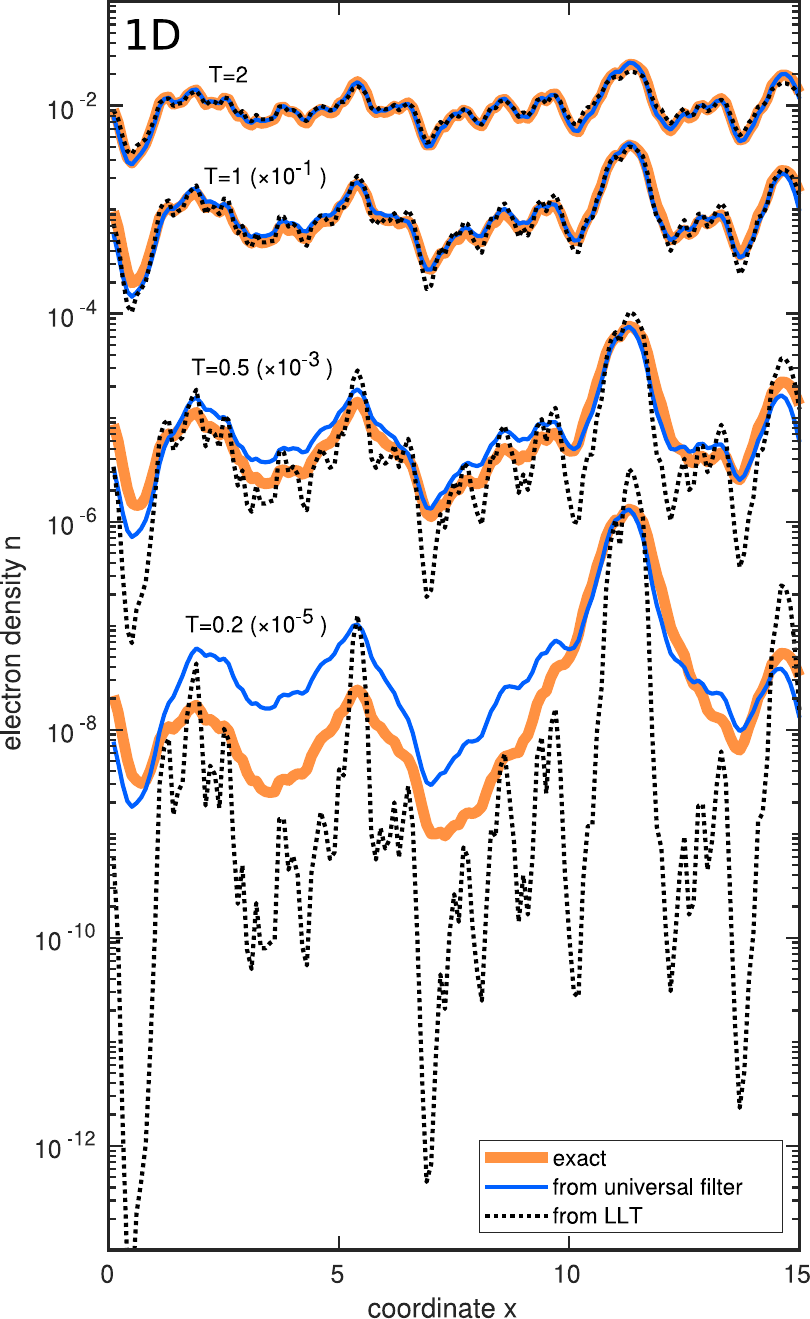}
	\caption{Comparison between the exact electron density
          $n(x,T)$
          (solid orange lines) and that obtained using
          the universal low-pass filter (ULF) function
          (solid blue lines) and the LLT (dashed black line)
          in a one-dimensional white-noise potential
          at different temperatures $T$. The parameters are the same as
          in Fig.~\ref{fig:algorithm1}, except for the temperature $T$.
          For clarity, the curves related to $T=0.2$, $T=0.5$ and $T=1$
          are multiplied by $10^{-5}$, $10^{-3}$ and $10^{-1}$, respectively,
          as indicated in the plot.
          The particle concentration is $n = 0.01$.\label{fig:1d_n_comparison}}
\end{figure}

\subsection{Temperature-dependent particle density}
\label{subsec:particledensity}

In this subsection, we first focus on the electron density in a one-dimensional setting.
The LLT is seen to work very well for temperatures $T\gtrsim T_0$
but the results deteriorate quickly for $T\lesssim 0.5T_0$.
Second, we investigate two dimensions where the bare LLT predicts too sharp
and pronounced maxima and minima even at $T=T_0$.

The LLT problem can be cured by a Gaussian broadening
of the random potential $V(\r)$ before the LLT is applied.
Such a broadening has been used by Piccardo et al.~\cite{LL2_2017}
and by Li et al.~\cite{LL3_2017}
in order to smoothen the rapidly changing distribution of atoms and to
obtain a continuous fluctuating potential.
Unfortunately, the optimal broadening parameter $\sigma(T)$
depends on temperature and is not known a-priori.

\subsubsection{Carriers on a chain}

In Sect.~\ref{subsec:univ_filter_Numerical tests},
we compared exact and approximate results
using the universal filtering in $d=1$ dimension.
Let us now compare these data with the LLT predictions for $n(x,T)$
at temperatures $T=0.2,0.5,1,2$ at $n=0.01$.

In Fig.~\ref{fig:1d_n_comparison} we plot the results of the LLT
along with the exact solution and with the ULF results copied from Fig.~\ref{fig:1d_n}.
Remarkably, the higher the temperature $T$, the better the agreement
between the three approaches.
Above $T=1$, the agreement can be considered as perfect with respect to both,
the number of extrema and the values $n(x,T)$.
At $T=0.5$, the LLT predicts much more extrema and a broader distribution for $n(x,T)$
in comparison with the exact solution and the ULF approach.
At $T=0.2$ and below, the disagreement between the LLT and the exact solution
increases drastically.

We conclude that the LLT is a reliable approach in $d=1$
at temperatures $T \gtrsim T_0$, though not at $T < T_0$.
Recall that $T_0 \equiv \varepsilon_0/k_B$,
where $\varepsilon_0$ is the energy scale of the disorder potential.
To give an example,
let us consider an InGaN multi-quantum-well solar-cell structures
designed for photo-carrier collection~\cite{WeisbuchNakamura2021}.
Using the estimate $\varepsilon_0 \approx 20$ meV from the energy slope
of the Urbach tail~\cite{WeisbuchNakamura2021},
we arrive at the estimate $T_0 \approx 230\, {\rm K}$
for the temperature below which the LLT approach
fails to describe the spatial distribution of the electron concentration.

\subsubsection{Carriers on a square lattice}
\label{subsec:carriers_on_a_square_lattice}

\begin{figure}
	\includegraphics[width=0.9\linewidth]{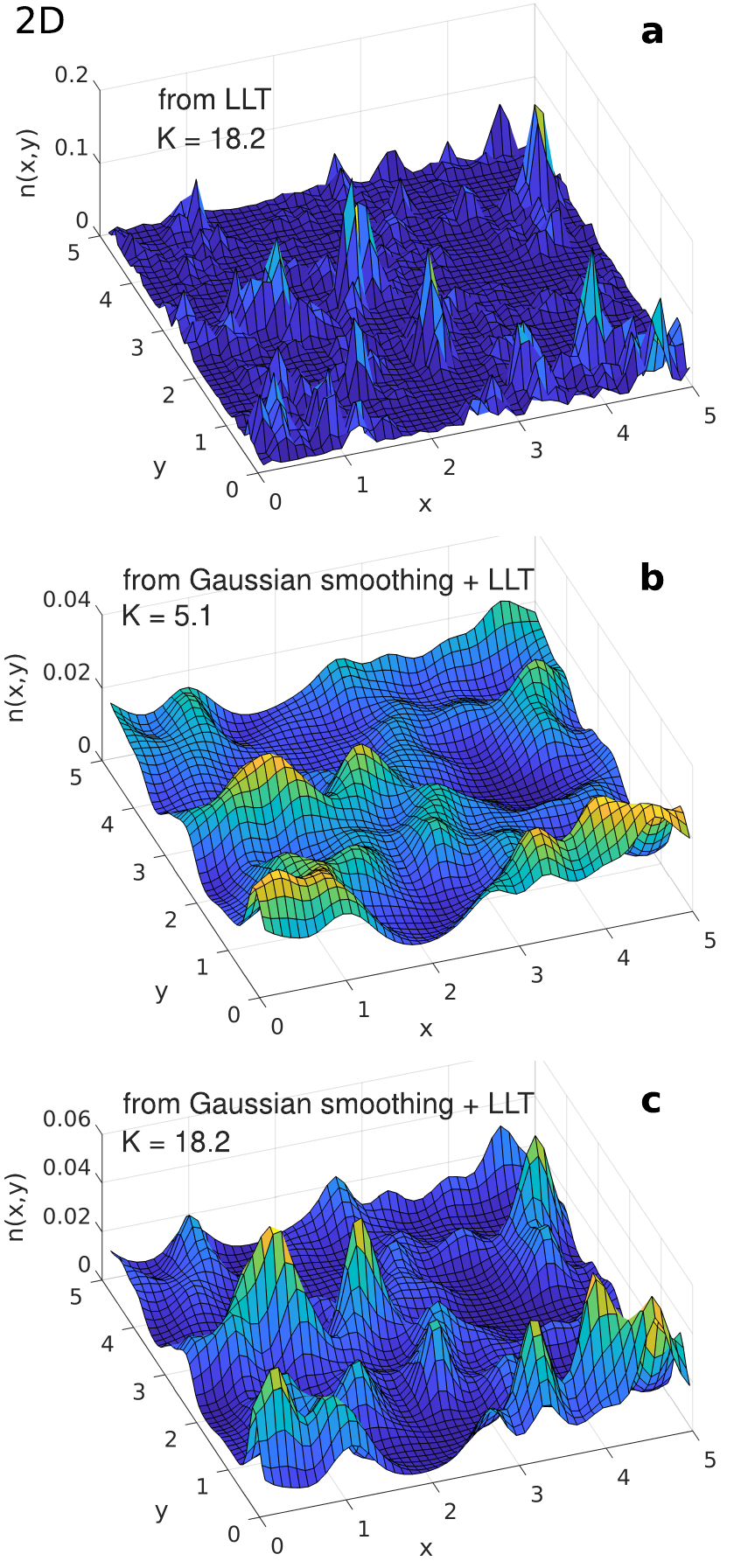}
	\caption{Electron density $n(x,y,T)$ obtained using the LLT approach
          with different parameters $\sigma$ in units $a$ and $K$ in units $\varepsilon_0$,
          as described in the text. (a) $\sigma = 0$, $K = 18.2$;
          (b) $\sigma = 1.5$, $K = 5.1$;
          (c) $\sigma = 1.5$, $K = 18.2$.
          Other parameters are the same as in
          Fig.~\protect\ref{fig:2d_n}.\label{fig:2d_n_LLT}}
\end{figure}

In Fig.~\ref{fig:2d_n_LLT}a,
we show the electron density $n(x,y,T)$ in $d=2$ at $T=T_0$, obtained
from the LLT with $K=18.2$ in units of $\varepsilon_0$
applied to the random potential $V(x,y)$ used for calculations of $n(x,y,T)$
in Fig~\ref{fig:2d_n}.
Apparently, the electron density provided by the LLT in Fig.~\ref{fig:2d_n_LLT}a
substantially deviates from $n(x,y,T)$ in Fig~\ref{fig:2d_n}.
The LLT predicts a series of narrow peaks in the electron density distribution
with amplitudes that are one order of magnitude larger than the values of $n(x,y,T)$
given by the exact solution.

In Fig.~\ref{fig:2d_n_LLT}b, we show the electron density $n(x,y,T)$ in $d=2$
at $T=T_0$, obtained by the application of the LLT to the random potential $V(x,y)$
smoothed by a Gaussian averaging.
The averaging is a convolution between $V(x,y)$ and a Gaussian function
$\exp[-(x^2+y^2)/(2\sigma^2)]$. Such a procedure has been previously used
by Piccardo et al.~\cite{LL2_2017} and by Li et al.~\cite{LL3_2017}
in order to smoothen the rapidly changing distribution of atoms and to
obtain a continuous fluctuating potential.
The smoothed potential allows a choice of the constant~$K$ in Eq.~(\ref{eq:LLT_K})
that is much smaller than that for the bare random potential $V(x,y)$
because the distribution of the potential values narrows after the Gaussian smoothing.
For the case of $V(x,y)$ used to generate Fig.~\ref{fig:2d_n_LLT}a,
the choice $K=5.1$ warrants a positive Hamiltonian for the smoothed potential
for which we choose $\sigma=1.5a$.

In order to preserve
the disorder effects on the electronic properties, the
length scale of the averaging must be smaller or comparable to
the typical scale of the effective potential fluctuations seen by
the carriers~\cite{LL2_2017,LL3_2017}. Piccardo et al.~\cite{LL2_2017}
and  Li et al.~\cite{LL3_2017} fixed the Gaussian broadening parameter to
a value of $\sigma = 2a_{\textrm{cat}} \approx 0.6$ nm, where $a_{\textrm{cat}}$
is the average distance between cations in GaN studied by Piccardo
et al.~\cite{LL2_2017}.
We varied the broadening parameter in order to achieve
the best agreement between the LLT result and the exact solution for $n(x,y,T)$
in $d=2$ at $T=T_0$. In Fig.~\ref{fig:2d_n_LLT}b, the electron density $n(x,y,T)$
is shown for the optimal value of the broadening parameter $\sigma = 1.5a$,
where $a$ is the distance between the grid points.

Apparently, it is possible to achieve the correct
electron density $n(x,y)$ in the framework of the LLT,
provided the optimal broadening  parameter for the Gaussian averaging
of the bare random potential $V(x,y)$ is known.
However, in lack of an exact solution, there is no recipe
in the framework of the LLT for how to get access to this optimal averaging parameter.
Further below, we shall give an heuristic estimate for $\sigma(T)$.

The question arises whether the difference in the spatial distributions
of the electron density in Fig.~\ref{fig:2d_n_LLT}a and in Fig.~\ref{fig:2d_n_LLT}b
is caused by the potential averaging itself,
or by the different choices of the constant $K$.
In order to resolve this question, we perform the calculations of $n(x,y,T)$
with the same smoothing of $V(x,y)$ as used for the data in Fig.~\ref{fig:2d_n_LLT}b
and the value $K=18.2$ as used for the data in Fig.~\ref{fig:2d_n_LLT}a.
The results are plotted in Fig.~\ref{fig:2d_n_LLT}c.
The distribution $n(x,y,T)$ in Fig.~\ref{fig:2d_n_LLT}c
drastically deviates from the correct result in Fig.~\ref{fig:2d_n}.
This underlines the importance of the constant~$K$ in Eq.~(\ref{eq:LLT_K})
for the predictions of the LLT approach.
While smoothing $V(x,y)$ with $\sigma = 1.5a$ and using $K = 5.1$
yields the correct distribution $n(x,y)$ for the given realization $V(x,y)$,
other choices for $\sigma$ and $K$ fail to reproduce the correct $n(x,y)$.

\subsection{Semiclassical transport}
\label{subsec:transport}

As a second application, we compare the predictions of RWF, ULF, and LLT
for semi-classical transport.

\subsubsection{Transport in one dimension}

First, we address the consequences of the various approaches
to the electron density $n(x,T)$
for the charge carrier mobility in one-dimensional disordered systems.
For simplicity, we assume a constant, spatially independent value $\mu_0$ for the local
mobility~\cite{IGZO2019}.

Since we have a series of local resistors, we address the local resistivity that is given by
\begin{equation}
\rho(x,T) = \frac{1}{e\mu_0 n(x,T)} \; .
\end{equation}
The sample average gives
\begin{equation}
  \bar{\rho}(T)\equiv
  \langle \rho(x,T) \rangle = \frac{1}{e\mu_0}\langle  n^{-1}(x,T)\rangle  \; .
  \label{eq:rhobar1}
\end{equation}
By definition of the macroscopic mobility we have
\begin{equation}
  \bar{\rho}(T)
  = \frac{1}{e\mu_{\rm eff}(T) \langle n(x,T)\rangle}\; .
 \label{eq:rhobar2}
\end{equation}
A comparison of eqs.~(\ref{eq:rhobar1}) and~(\ref{eq:rhobar2})
provides the following compact result for the macroscopic mobility $\mu_{\rm eff}(T)$
\begin{equation}
  \mu_{\rm eff}(T) = \frac{\mu_0}{\langle n(x,T)\rangle \langle n^{-1}(x,T)\rangle} \; .
  \label{eq:mu}
\end{equation}
In one dimension, the macroscopic mobility follows
from the average of the local particle density and of its inverse.

\begin{figure}[t]
	\includegraphics[width=\linewidth]{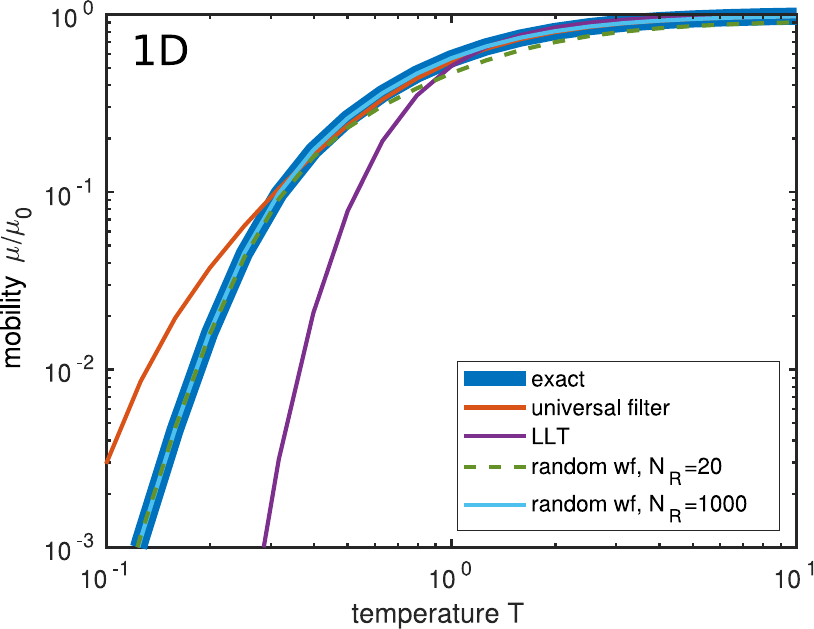}
	\caption{Temperature-dependent charge carrier
          mobility in one dimension. We compare exact results with those
          from the RWF, ULF, and LLT approximations to calculate the average of the local
          density $n(x,T)$ and of its inverse. $N_{\rm R}$ denotes the number of realizations
          used for averaging in the RWF algorithm.\label{fig:1d_mobility}}
\end{figure}

In Fig.~\ref{fig:1d_mobility} we show the exact result for
the temperature-dependent carrier mobility $\mu_{\rm eff}(T)$
in comparison with those from the various approximate approaches.
At $T \geq T_0$ all results practically coincide.
The RWF is the only approximate approach
that reliably reproduces $\mu_{\rm eff}(T)$ down to
temperatures, $T \approx 0.1 T_0$.
The results of the ULF approach start to deviate from the exact solution
around $T = 0.25 T_0$ but those of the LLT method begin to fail
already at $T = 0.8 T_0$, showing that the ULF is superior to the LLT
in accuracy, at smaller computational cost. Note, however,
that for very small temperatures, $T\lesssim 0.2T_0$, the ULF is also
inadequate for the mobility in one dimensional systems.

\subsubsection{Transport on a square lattice from percolation theory}

In dimensions $d\geq 2$, the calculation of $\mu_{\rm eff}(T)$
poses a percolation problem~\cite{Baranovskii2006Book}.
In essence, the percolation approach requires to
find the smallest value $\varsigma_{\rm c}(T)$ that provides a connected path via
areas with local conductivity $\varsigma(x,y,T) \geq \varsigma_{\rm c}(T)$.
The value $\varsigma_{\rm c}(T)$ is to be considered
as the macroscopic conductivity of the system~\cite{Baranovskii2006Book}
that characterizes the long-range charge transport.

\begin{figure}[b]
  \includegraphics[width=\linewidth]{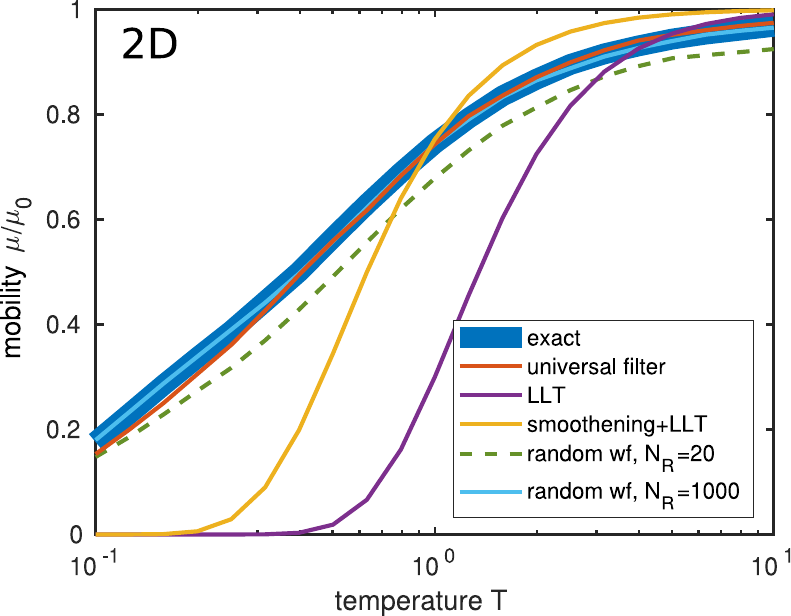}
	\caption{Temperature-dependent charge carrier
          mobility in two dimensions. We compare exact results from percolation theory
          with those
          from the RWF, ULF, and LLT approximations to calculate the average of the local
          density $n(x,T)$ and of its median. $N_{\rm R}$ denotes the number of realizations
          used for averaging in the RWF algorithm.
          LLT results with and without Gaussian smoothing.\label{fig:2d_mobility}}
\end{figure}

Percolation theory predicts that in $d=2$ the area corresponding
to the inequality $\varsigma(x,y,T) \geq \varsigma_{\rm c}(T)$
is exactly one half of a total system area~\cite{Shklovskii1984}.
Herewith, $\varsigma_{\rm c}(T)$ is the median of the distribution $\varsigma(x,y,T)$.
The local conductivity is determined by the local electron concentration $n(x,y,T)$,
\begin{equation}
  \varsigma(x,y,T) = e \mu_0 n(x,y,T)\; ,
\end{equation}
while the macroscopic conductivity $\varsigma_{\rm c}(T)$
is determined by the average concentration $\langle n(x,y,T)\rangle$,
\begin{equation}
  \varsigma_{\rm c}(T) = e \mu_{\rm eff}(T) \langle n(x,y,T)\rangle\; .
\end{equation}
The latter equation yields the relation
\begin{equation}
\mu_{\rm eff}(T) = \mu_0 \frac{\overline{n(x,y,T)}}{\langle n(x,y,T)\rangle} \; ,
  \label{eq:mu_2d}
\end{equation}
where $\overline{n(x,y,T)}$ denotes the median of the distribution $n(x,y,T)$.

In Fig.~\ref{fig:2d_mobility} we show the results for the temperature-dependent
carrier mobility $\mu_{\rm eff}(T)$ in two dimensions given by eq.~(\ref{eq:mu_2d})
as obtained from the different schemes for calculating the average and the median
of $n(x,T)$.
The RWF and the ULF provide reliable results down to $T=0.1 T_0$,
the results of the LLT deviate from the exact data already at $T=5T_0$.
The agreement becomes slightly better when a Gaussian averaging is applied to the
disorder potential $V(x,y)$ before the LLT is applied, see
Sect.~\ref{subsec:carriers_on_a_square_lattice}.
When the Gaussian broadening $\sigma = 1.5a$
and the constant $K = 5.1 \varepsilon_0$ are used in eq.~(\ref{eq:LLT_K})
so that the distribution $n(x,y)$ is reproduced from LLT at $T=T_0$
as seen in Sect.~\ref{subsec:carriers_on_a_square_lattice},
also the value of $\mu_{\rm eff}(T)/\mu_0$ calculated via eq.~(\ref{eq:mu_2d})
is reproduced at $T=T_0$ but not at any other temperature.
This strengthens the conclusion that the choice of the Gaussian broadening
parameter $\sigma$ and the choice of the constant $K$
in the LLT equation (\ref{eq:LLT_K}) must depend on temperature
to bring the result in agreement with the exact solution.

Heuristically, we find that the choice $\sigma(T) \sim T^{-0.75}$
provides acceptable results in two dimensions for the mobility $\mu_{\rm eff}(T)$.
After smoothing, we set $K=-1.05 \tilde{V}_{\rm min}$, where $\tilde{V}(x,y)$
is the random potential after Gaussian smoothing with its
absolute minimum $\tilde{V}_{\rm min}$.

\section{Discussion and conclusions}
\label{sec:conclusions}

The theoretical description of the opto-electronic properties of disordered media
requires an accurate knowledge of the space-dependent and temperature-dependent
charge carrier distribution $n(\r,T)$ in the presence of a random potential.
Two methods are currently available to determine $n(\r,T)$:
(i) solving the Schr\"odinger equation
and (ii) utilizing the localization
landscape theory (LLT)~\cite{Arnold2016,LL1_2017,LL2_2017,LL3_2017}.
While exact, the complete solution of the Schr\"odinger equation is extremely demanding
with respect to computational time and computer memory.
It is hardly affordable for applications to realistically large, chemically complex systems.
As exemplified in Sect.~\ref{sec:ComparisonwithLLT}, the LLT
is valid only in certain cases with unspecific limits.
These limits can be revealed only by comparison with the exact solution
of the Schr\"odinger equation for particles in a random potential $V(\r)$.

In this work, we propose two novel theoretical tools, the random-wave-function
(RWF) algorithm and the universal low-pass filter (ULF) approach,
that permit an approximate calculation of $n(\r,T)$
without solving the Schr\"odinger equation.
In comparison, both methods require less computational resources than
the complete solution of the Schr\"odinger equation, and
both have a better accuracy and a broader range of applicability than the LLT.

As shown in this work, the RWF approach is as accurate and generally applicable to
non-degenerate electron systems as solving the Schr\"odinger equation.
However, for practical applications at temperatures $T\gtrsim 0.1 T_0$,
the RWF is much less demanding
with respect to computational resources than solving the Schr\"odinger problem.
Nevertheless, even the RWF becomes computationally too costly for
mesoscopically large three-dimensional systems at low temperatures.
For example, the calculations of $n(\r,T)$ depicted in Fig.~\ref{fig:3d_n}
consumed 1.5~hours on a PC in the case of the RWF approach
but only 0.15~seconds in the case of the ULF scheme.

The ULF approach employs the temperature-de\-pen\-dent effective potential $W(\r,T)$.
In this respect it is similar to the LLT, which also relies on a quasi-classical potential
that replaces the random potential $V(\r)$ . However,
the effective potential $W(\r,T)$ in the ULF approach depends on temperature,
while the effective potential in LLT is temperature-independent.
In Figs.~\ref{fig:W_statisticsagain} and \ref{fig:1d_W}, it is clearly seen that the effective potential responsible
for the distribution $n(\r,T)$ strongly depends on temperature to reproduce
the exact solution. Sect.~\ref{sec:ComparisonwithLLT}
provides evidence that ignoring the $T$-dependence of the effective potential $W(\r,T)$
prevents an accurate calculation of the electron distribution $n(\r,T)$.
In the ULF, a universal low-pass filter is applied to the potential $V(\r)$
via eq.~(\ref{eq:W-via-V}). We derive this linear filter analytically
using high-temperature perturbation theory.
The filter is $T$-dependent, while $T$ does not enter Eq.~(\ref{eq:LLT}) of the LLT.

The numerical tests performed in Sect.~\ref{sec:ComparisonwithLLT} show
that the range of applicability for the ULF approach is much broader
than that for the LLT. Furthermore, the ULF scheme requires only
Fast Fourier Transformation which is computationally much less demanding
than solving the LLT eq.~(\ref{eq:LLT}) so that mesoscopically large three-dimensional
disordered systems appear within reach for temperatures $T\gtrsim 0.5T_0$.

The space- and temperature-dependent electron distribution $n(\r, T)$
is the key ingredient for the theoretical treatment of charge transport. In our work
it is shown that RWF and ULF provide a solid basis for the theoretical treatment
of charge transport in disordered media.

In this paper, the calculation of the equilibrium carrier density in the Boltzmann approximation is considered for a disordered potential as a function of temperature. However, in practice, there are many problems where the carrier distribution at low temperature is not in equilibrium. Although neither RWF, nor ULF can be applied to non-equilibrium problems straightforwardly, there is a perspective to possible modification of RWF in order to make it applicable to non-equilibrium cases. If the number of realizations $N_R$ decreases, the calculated occupation of energy levels by electrons gradually deviates from their equilibrium values. In principle, these deviations may mimic the non-equilibrium distributions, and one could interpret the green dashed lines in Figs.~\ref{fig:algorithm1} and~\ref{fig:algorithm4} as corresponding to the non-equilibrium electron densities. However, two problems arise in such interpretation of the calculated results. First, the deviations from the equilibrium occupation should be more pronounced for the low-energy states than for the high-energy levels. Some modifications of RWF should be performed to fulfill this condition. Second, the occupations of energy levels in RWF are described by $\chi^2$ distribution, while in reality another statistics holds. For instance, the occupations should obey Poisson statistics in the approximation of non-interacting electrons. More research is necessary to overcome these obstacles.

\begin{acknowledgments}
A.N. thanks the Faculty of Physics of the Philipps Universit\"at Marburg
for the kind hospitality during his research stay. S.D.B. and K.M. acknowledge financial
support by the Deutsche
Forschungsgemeinschaft (Research Training Group ``TIDE'', RTG2591)
as well as by the key profile area ``Quantum Matter and Materials
(QM2)'' at the University of Cologne. K.M. further acknowledges support
by the DFG through the project ASTRAL (ME1246-42).
\end{acknowledgments}

\appendix

\section{Accuracy of the random-wave-function algorithm}
\label{sec:appendix-accuracy}

The aim of this section is to derive expression~(\ref{eq:Delta2-expected})
for the expected value of $[\Delta(\r)]^2$.
Dividing the numerator and the denominator by $e^{\beta\mu}$
in eq~(\ref{eq:Delta-def}), and taking eqs.~(\ref{eq:n-via-tilde-n})
and (\ref{eq:n-from-algorithm}) into account, we obtain
\begin{equation}
  \Delta(\r) = \frac{ N_{\rm R}^{-1} \sum_{\rm R} [ \tilde{n}_{\rm R} (\r)
      - \tilde{n}(\r) ] }{\tilde{n}(\r) } \; .
  \label{eq:app-Delta}
\end{equation}

As shown in Section~\ref{subsec:analysis},
\begin{equation}
  \left\langle \tilde{n}_{\rm R}(\r) \right\rangle = \tilde{n}(\r) \; ,
  \label{eq:app-aver-tilde-n}
\end{equation}
where angle brackets stand for the expected value. Substituting this
result into eq.~(\ref{eq:app-Delta}), one can ensure that
\begin{equation}
  \left\langle  \Delta(\r) \right\rangle = 0 \; .
  \label{eq:aver-Delta-is-zero}
\end{equation}
Hence,
\begin{eqnarray}
    \left\langle  [\Delta(\r)]^2 \right\rangle &=& \text{var}\,\Delta(\r) \nonumber \\
    &=& \frac{ \text{var}\,\{ \sum_{\rm R}
      [ \tilde{n}_{\rm R} (\r) - \tilde{n}(\r) ] \} }{N_{\rm R}^2 \,
      [\tilde{n}(\r)]^2 } \nonumber \\
    &=& \frac{ \sum_{\rm R} \text{var}\,\{
      [ \tilde{n}_{\rm R} (\r) - \tilde{n}(\r) ] \} }{
      N_{\rm R}^2 \, [\tilde{n}(\r)]^2 } \nonumber \\
    &=& \frac{ \text{var}\,\{  [ \tilde{n}_{\rm R} (\r) - \tilde{n}(\r) ] \} }{
      N_{\rm R} \, [\tilde{n}(\r)]^2 } \nonumber \\
    &=& \frac{  \left\langle  [\tilde{n}_{\rm R} (\r)]^2 \right\rangle
      - [\tilde{n}(\r)]^2 }{N_{\rm R} \, [\tilde{n}(\r)]^2 } \; ,
  \label{eq:aver-Delta2}
\end{eqnarray}
where symbol `var' denotes variance, and we use the fact that variance of the
sum is equal to the sum of variances of independent variables.

Then we recall that $\tilde{n}_{\rm R} (\r)$ is a square of random quantity
$\gamma_{\rm R} (\r) \equiv \sqrt2 \psi_{\rm R} (\r)$ according to
eq.~(\ref{eq:psi2-via-tilde-n}). The latter quantity is, by construction,
a linear combination of Gaussian random variables $c_{i,R}^{(0)}$,
and therefore has a Gaussian distribution with expected value zero.
The expected value of $\tilde{n}_{\rm R} (\r)$ is the second moment of
$\gamma_{\rm R} (\r)$, and the expected value of $[\tilde{n}_{\rm R} (\r)]^2$
is the fourth moment of $\gamma_{\rm R} (\r)$. Due to Wick's probability
theorem, the fourth moment of a normally distributed random variable is
three times larger than the square of its second moment. Hence,
\begin{equation}
  \left\langle  [\tilde{n}_{\rm R} (\r)]^2 \right\rangle = 3 [\tilde{n}(\r)]^2 .
  \label{eq:fourth-moment-to-second}
\end{equation}
Finally, substitution of eq.~(\ref{eq:fourth-moment-to-second})
into eq.~(\ref{eq:aver-Delta2}) gives
\begin{equation}
  \left\langle  [\Delta(\r)]^2 \right\rangle = \frac{2}{N_{\rm R}} \; .
  \label{eq:aver-Delta2-result}
\end{equation}

\section{Constant parameter at small discretization}
\label{app:B}

In this appendix we show how to derive the scaling form~(\ref{eq:CpropSformula})
for the constant parameter $C(T)$.

We start from eq.~(\ref{eq:n-via-W})
that expresses the electron density $n(\r,T)$ through the effective potential $W(\r,T)$,
\begin{eqnarray}
  n(\r,T) &=& N_c \, e^{\beta[\mu-W(\r,T)]} \nonumber\\
  &=& N_c e^{\beta\mu}
  \left[ 1 - \beta W(\r,T) + \frac{\beta^2}{2} W^2(\r,T) - \ldots \right] \; . \nonumber \\
\label{eq:app-n-via-W}
\end{eqnarray}
Since we consider the potential $V(\r)$ as a small perturbation, we
expand $n(\r,T)$ and $W(\r,T)$ in a power series,
\begin{eqnarray}
  n(\r,T) &=& n_0(T)+n_1(\r,T)+n_2(\r,T) + \mathcal O(V^3) \; ,\quad
      \label{eq:app-n1-n2}\\
  W(\r,T) &=& W_1(\r,T) + W_2(\r,T) + \mathcal O(V^3) \; , \quad
    \label{eq:app-W1-W2}
\end{eqnarray}
where $n_0(T)= N_c \exp(\beta\mu)$. Here,
$n_1(\r,T)$ and $W_1(\r,T)$ are linear in $V$,
and $n_2(\r,T)$ and $W_2(\r,T)$ are quadratic in $V$.
Substituting eq.~(\ref{eq:app-n1-n2})
into eq.~(\ref{eq:app-n-via-W}), and collecting the terms up to quadratic order
in $V$, we obtain
\begin{equation}
  n_2(\r,T) = N_c \, e^{\beta\mu}
  \left[ \frac{\beta^2}{2} W_1^2(\r,T) - \beta W_2(\r,T) \right] \; ,
  \label{eq:app-n2-via-W}
\end{equation}
where the filter function defines the linear relation between the random potential
and the effective potential,
\begin{equation}
  W_1(\r,T) = \int {\rm d} \r' \Gamma(\r',T) V(\r+\r') \; ,
    \label{eq:app-W1-via-Gamma}
\end{equation}
see eqs.~(\ref{eq:W-via-V}) and~(\ref{eq:Gamma-n-perturb-only-kinetic-1}).

Eq.~(\ref{eq:C-value}) shows that $C(T)=\langle W(\r) \rangle
- \langle V(\r) \rangle$. The odd powers of the white-noise potential $V$
vanish after averaging, hence
\begin{equation}
  C(T) = \langle W_2(\r,T) \rangle + \mathcal O(V^4) \; .
  \label{eq:app-C-via-W}
\end{equation}
We substitute $W_2(\r,T)$ using eq.~(\ref{eq:app-n2-via-W}) into
eq.~(\ref{eq:app-C-via-W}) and neglect the residual term $\mathcal O(V^4)$
to obtain
\begin{equation}
  C(T) = C^{(1)}(T) + C^{(2)}(T)
  \label{eq:app-C-via-C12}
\end{equation}
with the two contributions
\begin{eqnarray}
  C^{(1)}(T) &=& \frac{\beta}{2} \, \langle W_1^2(\r,T) \rangle \; , \nonumber \\
    C^{(2)}(T) &=& - \, \frac{\langle n_2(\r,T) \rangle}{\beta N_c \, e^{\beta\mu}} \; .
  \label{eq:app-C12-def}
\end{eqnarray}

\subsection{First contribution}

The mean value of $W_1^2(\r,T)$ in eq.~(\ref{eq:app-C12-def})
depends on the statistics of the random potential $V(\r)$.
For a white-noise potential, eq.~(\ref{eq:white-noise-def}) leads to
\begin{equation}
  \langle W_1^2(\r,T) \rangle = S \int \d\r' \, [\Gamma(\r',T)]^2 \; ,
  \label{eq:app-aver-W1-via-Gamma}
\end{equation}
which is indeed independent of~$\r$.
Here, the parameter $S$ characterizes the strength of the potential fluctuations.
The Fourier image of the filter function,
\begin{equation}
  \hat{\Gamma}(\k) = \int \d\r \, \Gamma(\r) \, e^{\rmi\k\cdot\r} \; ,
  \label{eq:app-Gamma-k-def}
\end{equation}
permits to rewrite eq.~(\ref{eq:app-aver-W1-via-Gamma}) using the Plancherel
identity with the result
\begin{equation}
  C^{(1)}(T) = \frac{\beta S}{2(2\pi)^d} \int \d\k \, [\hat\Gamma(\k)]^2 \; .
  \label{eq:app-C1-via-Gamma}
\end{equation}
Here, $d=1,2,3$ is the spacial dimension.
Using Eq.~(\ref{eq:hat-Gamma-via-hat-Gamma-uni}),
we express $\hat{\Gamma}(\k)$ via the
universal function $\hat{\Gamma}_{\text{uni}}(k)$,
which is defined by eq.~(\ref{eq:hat-Gamma-uni}).
The resulting coefficients $C^{(1)}(T)$ are
\begin{eqnarray}
  C^{(1)}_{1d} (T)
  &=& \frac{\beta S}{4\pi\lambdabar}
  \int_{-\infty}^{\infty}  \left[ \frac{\sqrt\pi}{k} e^{-k^2/4} \, \text{erfi}(k/2) \right]^2\d k
  \nonumber \\
&  \approx& 0.260 \, \frac{\beta S}{\lambdabar}
  \label{eq:app-C1-1d-result}
\end{eqnarray}
in $d=1$ dimension,
\begin{eqnarray}
  C^{(1)}_{2d}(T)
  &=& \frac{\beta S}{2(2\pi\lambdabar)^2}
  \int_0^{\infty} \!\! 2\pi k \left[ \frac{\sqrt\pi}{k} e^{-k^2/4}
    \text{erfi}(k/2) \right]^2 \d k \nonumber \\
 & \approx& 0.146 \, \frac{\beta S}{\lambdabar^2}
  \label{eq:app-C1-2d-result}
\end{eqnarray}
in $d=2$ dimensions,
and
\begin{eqnarray}
  C^{(1)}_{3d} (T)
  &=& \frac{\beta S}{2(2\pi\lambdabar)^3} \int_0^{\infty}\!\!
  4\pi k^2 \left[ \frac{\sqrt\pi}{k} e^{-k^2/4} \text{erfi}(k/2) \right]^2 \d k
  \nonumber \\
  &\approx& 0.0971 \, \frac{\beta S}{\lambdabar^3}
  \label{eq:app-C1-3d-result}
\end{eqnarray}
in $d=3$ dimensions.

\subsection{Second contribution}

To  calculate $C^{(2)}(T)$,
we need an expression for the second-order correction $n_2(\r,T)$
to the electron density $n(\r,T)$.
We start from eq.~(\ref{eq:n-via-exp-H})
that expresses the electron density through the
one-particle Hamiltonian $\hat{H}$,
\begin{equation}
  n(\r,T) = 2e^{\beta\mu} \langle \r | \exp(-\beta\hat{H}) | \r \rangle\; .
\end{equation}
Let us consider the kinetic-energy operator $\hat{T}$
as the non-perturbed Hamiltonian, and the potential-energy operator
\begin{equation}
  \hat{V} = \int \d\r  \, V(\r) \, |\r\rangle \langle\r|
  \label{eq:app-V-def}
\end{equation}
as a small perturbation. Then, the perturbation expansion up to the second order yields
\begin{eqnarray}
  n(\r,T) &=& n_0(T)+ n_1(\r,T)+ n_2(\r,T) \; , \nonumber \\
  n_0(T) &=& 2 e^{\beta\mu} \langle\r| e^{-\beta \hat{T}} |\r\rangle \nonumber\; ,\\
  n_1(\r,T) &=&- 2 e^{\beta\mu} \int_0^\beta \d\xi \, \langle\r|
  e^{-\xi \hat{T}} \hat{V} e^{(\xi-\beta) \hat{T}} |\r\rangle \nonumber\; ,\\
  n_2(\r,T)&=& 2 e^{\beta\mu} \int_0^\beta \d\xi_1 \int_{\xi_1}^\beta \d\xi_2 \nonumber \\
&&   \times \langle\r| e^{-\xi_1 \hat{T}} \hat{V}e^{(\xi_1-\xi_2) \hat{T}}
  \hat{V}e^{(\xi_2-\beta) \hat{T}} |\r\rangle \; .\quad
    \label{eq:app-n-perturb-expansion}
\end{eqnarray}
As our next step, we insert the expression for $\hat{V}$ from eq.~(\ref{eq:app-V-def})
into eq.~(\ref{eq:app-n-perturb-expansion}) to see that
\begin{equation}
  n_2(\r,T) = \iint \d\r_1 \, \d\r_2 \, \Gamma_{n}^{(2)}(\r_1, \r_2,T)
  V(\r+\r_1) \, V(\r+\r_2) \; ,
  \label{eq:app-n2-via-Gamma-n2}
\end{equation}
where the kernel $\Gamma_{n}^{(2)}(\r_1, \r_2,T)$ is defined by
\begin{eqnarray}
  \Gamma_{n}^{(2)}(\r_1, \r_2,T)
  &=& 2 e^{\beta\mu} \int_0^\beta \d\xi_1 \int_{\xi_1}^\beta \d\xi_2 \,
  \langle \veczero| e^{-\xi_1 \hat{T}} |\r_1\rangle \nonumber \\
  &&\times
  \langle\r_1| e^{(\xi_1-\xi_2) \hat{T}} |\r_2\rangle
  \langle\r_2| e^{(\xi_2-\beta) \hat{T}} |\veczero\rangle \; .\nonumber \\
  \label{eq:app-Gamma-n2-perturb}
\end{eqnarray}
The expression~(\ref{eq:app-C12-def}) for $C^{(2)}(T)$
contains the mean value of $n_2(\r,T)$.
Using eq.~(\ref{eq:app-n2-via-Gamma-n2}) and the statistics of the white-noise
potential $V(\r)$ from eq.~(\ref{eq:white-noise-def}) we
obtain the average value in the form
\begin{equation}
  \langle n_2(\r,T) \rangle = S \int \d\r' \Gamma_{n}^{(2)}(\r', \r',T) \; ,
  \label{eq:app-aver-n2-via-Gamma-n2}
\end{equation}
which is independent of~$\r$.
To calculate the integral on the right-hand side,
we represent the kinetic-energy operator $\hat{T}$ in Fourier space as
\begin{equation}
  \hat{T} = \frac{\Omega}{(2\pi)^d}
  \int \d\k  \, \frac{\hbar^2 k^2}{2m} \, |\k\rangle \langle\k| \;,
  \label{eq:app-T-def}
\end{equation}
and use eq.~(\ref{eq:k-r}) for the plane-wave matrix elements $\langle\r|\k\rangle$.
Then eq.~(\ref{eq:app-Gamma-n2-perturb}) leads to
\begin{eqnarray}
  \int \d\r \, \Gamma_{n}^{(2)}(\r,\r,T) &=&
  \frac{2e^{\beta\mu}}{(2\pi)^{3d}} \int_0^\beta \d\xi_1 \int_{\xi_1}^\beta \d\xi_2
   \label{eq:app-int-Gamma-n2-1}\\
  && \iiint
  \d\k_1 \, \d\k_2 \, \d\k_3 \int \d\r
  \nonumber \\
&&
  \times  e^{-\hbar^2\xi_1k_1^2/(2m) }
  e^{\hbar^2(\xi_1-\xi_2)k_2^2/(2m) } \nonumber \\
&& \times   e^{\hbar^2(\xi_2-\beta)k_3^2/(2m) } e^{\rmi (\k_3-\k_1)\cdot\r}\; .
 \nonumber
\end{eqnarray}
Here, the integration over $\r$ is straightforward,
\begin{equation}
  \int \d\r  \, e^{\rmi(\k_3-\k_1)\cdot\r} = (2\pi)^d \, \delta(\k_3-\k_1) \; ,
  \label{eq:app-integral-delta}
\end{equation}
which allows us to perform the integral over $\k_3$.
Thence,
\begin{eqnarray}
  \int \d\r \, \Gamma_{n}^{(2)}(\r,\r,T)
  &=& \frac{2e^{\beta\mu}}{(2\pi)^{2d}}
  \int_0^\beta \d\xi_1 \int_{\xi_1}^\beta \d\xi_2 \nonumber \\
  &&
  \times
  \left( \int \d\k_1 \, e^{-(\beta-\xi_2+\xi_1)\hbar^2 k_1^2/2m} \right)  \nonumber \\
&&  \times \left( \int \d\k_2 \, e^{-(\xi_2-\xi_1)\hbar^2 k_2^2/2m} \right) \nonumber \\
  & = &
  \frac{2e^{\beta\mu}}{(2\pi)^{2d}} \int_0^\beta \d\xi_1 \int_{\xi_1}^\beta \d\xi_2
  \nonumber \\
  && \hphantom{\frac{2e^{\beta\mu}}{(2\pi)^{2d}}}\times
  \left( \frac{2\pi m}{(\beta-\xi_2+\xi_1)\hbar^2} \right)^{d/2}  \nonumber \\
  && \hphantom{\frac{2e^{\beta\mu}}{(2\pi)^{2d}}}
  \times   \left( \frac{2\pi m}{(\xi_2-\xi_1)\hbar^2} \right)^{d/2} .\quad
  \label{eq:app-int-Gamma-n2-2}
\end{eqnarray}
In the latter expression, the parameters $\xi_1$ and $\xi_2$
appear only in the combination $\xi = \xi_2-\xi_1$.
To further simplify the integrals, we note that, for any function $\phi(\xi)$,
\begin{equation}
  \int_0^\beta \d\xi_1 \int_{\xi_1}^\beta \d\xi_2 \, \phi(\xi_2-\xi_1)
  = \int_0^\beta \d\xi \, (\beta-\xi) \, \phi(\xi)
  \label{eq:app-int-xi12-via-int-xi}
\end{equation}
holds. Hence,
\begin{eqnarray}
  \int \d\r \, \Gamma_{n}^{(2)}(\r,\r,T)
  &=& 2e^{\beta\mu} \left(\frac{m}{2\pi\hbar^2}\right)^d \nonumber \\
&& \times   \int_0^{\beta} \d\xi \; \xi^{-d/2} (\beta-\xi)^{1-d/2} .\qquad
  \label{eq:app-int-Gamma-n2-3}
\end{eqnarray}
We combine eqs.~(\ref{eq:app-C12-def}),
(\ref{eq:app-aver-n2-via-Gamma-n2}), and~(\ref{eq:app-int-Gamma-n2-3}),
and express $N_c$ from eq.~(\ref{eq:Nc-def}) and $\lambdabar$ from
eq.~(\ref{eq:lambda-de-Broglie}) to arrive at a simple formula
for the constant $C^{(2)}(T)$,
\begin{equation}
  C^{(2)}(T) = - \frac{S \beta^{d-1}}{(2\sqrt\pi\lambdabar)^d}
  \int_0^\beta \d\xi \; \xi^{-d/2} (\beta-\xi)^{1-d/2} \; .
  \label{eq:app-C2-result}
\end{equation}
In $d=1$ dimension, this formula is readily evaluated,
\begin{equation}
  C^{(2)}_{1d}(T) = - \frac{\sqrt\pi}{4} \; \frac{S\beta}{\lambdabar}
  \label{eq:app-C2-1d-result}
\end{equation}
because the integral converges.

However, the integral in eq.~(\ref{eq:app-C2-result}) diverges
in two and three dimensions. In particular,
\begin{equation}
  C^{(2)}_{2d}(T) = - \frac{S\beta}{4\pi\lambdabar^2} \int_0^\beta \frac{\d\xi}{\xi} \, .
  \label{eq:app-C2-2d-1}
\end{equation}
This divergence stems from the divergent integral
\begin{equation}
I(\xi)=  \int \d\k \exp\left( -\xi \frac{\hbar^2 |\k|^2}{2m} \right) \;,
  \label{eq:app-xi-integral}
\end{equation}
that appears in eq.~(\ref{eq:app-int-Gamma-n2-2}).
This integral becomes infinitely large in the limit $\xi\to 0$ when
the integration is performed over the whole $\k$-space.
However, in a discretized model, the value of $|\k|$ cannot be larger
than $\simeq a^{-1}$, where $a$ is the grid lattice constant.
This shows that the lower limit of the integral in eq.~(\ref{eq:app-C2-2d-1})
should be set to
\begin{equation}
  \xi_{\text{min}} \simeq \frac{1}{\epsilon_{\text{max}}}
  \simeq \frac{ma^2}{\hbar^2} \simeq \frac{a^2\beta}{\lambdabar^2} \; .
  \label{eq:app-xi-min}
\end{equation}
Therefore, in two dimensions we find
\begin{equation}
  C^{(2)}_{2d}(T) = \frac{S\beta}{2\pi\lambdabar^2}
 \ln\frac{\tilde{\mu}_2 a}{\lambdabar}  \; ,
  \label{eq:app-C2-2d-result}
\end{equation}
where $\tilde{\mu}_2$ arises from the unknown proportionality factor
between the left-hand side and the right-hand side of eq.~(\ref{eq:app-xi-min}).
The factor $\mu_2$ may depend also on the type of the discretization
grid (square, hexagonal, etc.).

Similarly, we replace the lower limit in eq.~(\ref{eq:app-C2-result})
by $\xi_{\text{min}}$
also in three dimensions,
\begin{equation}
  C^{(2)}_{3d}(T) =
  - \frac{S\beta}{(4\pi)^{3/2}\lambdabar^3}
  \left[ 2\sqrt\frac{\beta}{\xi_{\text{min}}}
    + \mathcal O\left( \frac{\xi_{\text{min}}}{\beta} \right)  \right] \; .
  \label{eq:app-C2-3d-1}
\end{equation}
Neglecting the last term in the square brackets and substituting
$\xi_{\text{min}}$ from eq.~(\ref{eq:app-xi-min}), we find
\begin{equation}
  C^{(2)}_{3d}(T) = - \mu_3 \frac{S\beta}{\lambdabar^2a} \; ,
  \label{eq:app-C2-3d-result}
\end{equation}
where the parameter $\mu_3$ again depends
on the undetermined factor in eq.~(\ref{eq:app-xi-min}).

\subsection{Summary}

Summing up the expressions for $C^{(1)}(T)$,
eqs.~(\ref{eq:app-C1-1d-result})--(\ref{eq:app-C1-3d-result}),
and for $C^{(2)}(T)$, eqs.~(\ref{eq:app-C2-1d-result}),
(\ref{eq:app-C2-2d-result}), and~(\ref{eq:app-C2-3d-result}),
we arrive at the following results for the coefficient $C(T)$ in $d=1,2,3$ dimensions,
\begin{eqnarray}
  \label{eq:app-C-1d}
  C_{1d}(T) &=& - 0.183 \; \frac{S}{k_{\text{B}}T\lambdabar} \; , \\
  \label{eq:app-C-2d}
  C_{2d}(T) &=&\frac{S}{k_{\text{B}}T\lambdabar^2}\; \frac{1}{2\pi}
  \ln\frac{\mu_2a}{\lambdabar} \; , \\
  \label{eq:app-C-3d}
  C_{3d}(T) &=& \frac{S}{k_{\text{B}}T\lambdabar^3}
  \left( -\mu_3\frac{\lambdabar}{a} + 0.0971  \right) \; .
\end{eqnarray}
The parameters $\mu_2$ and $\mu_3$
in these formulas are lattice specific, and can be found by fitting of the numerical data.
As shown in the main text, $\mu_2 \approx 0.30$ for the square lattice,
and $\mu_3 \approx 0.255$ for the three-dimensional cubic lattice.
Thus eq.~(\ref{eq:CpropSformula}) is shown to hold within second-order
perturbation theory for a white-noise random potential.

\bibliography{LL_filter}

\end{document}